\documentclass{elsarticle}

\usepackage{hyperref}

\usepackage{lipsum}
\usepackage{amsmath}
\usepackage{amssymb}
\usepackage{booktabs}
\usepackage[table]{xcolor}
\usepackage{multirow}
\usepackage[caption=false,font=footnotesize]{subfig}

\makeatletter
\def\input@path{{./tables/}}
\makeatother

\newcommand{\Sc}{\mathcal{S}}
\newcommand{\degC}{$^{\circ}$C}

\graphicspath{{figures/}}

\begin{document}

\begin{frontmatter}

\title{Impacts of Heat Decarbonisation on System Adequacy\\considering Increased Meteorological Sensitivity}
\author[NU]{Matthew Deakin\corref{mycorrespondingauthor}}
\author[RU]{Hannah Bloomfield}
\author[NU]{David Greenwood}
\author[DU]{Sarah Sheehy}
\author[NU]{Sara Walker}
\author[BU]{Phil C. Taylor}

\address[NU]{Newcastle University, Newcastle-upon-Tyne, UK}
\address[RU]{University of Reading, Reading, UK}
\address[DU]{Durham University, Durham, UK}
\address[BU]{University of Bristol, Bristol, UK}

\begin{abstract}
This paper explores the impacts of decarbonisation of heat on demand and subsequently on the generation capacity required to secure against system adequacy standards. Gas demand is explored as a proxy variable for modelling the electrification of heating demand in existing housing stock, with a focus on impacts on timescales of capacity markets (up to four years ahead). The work considers the systemic changes that electrification of heating could introduce, including biases that could be introduced if legacy modelling approaches continue to prevail. Covariates from gas and electrical regression models are combined to form a novel, time-collapsed system model, with demand-weather sensitivities determined using lasso-regularized linear regression. It is shown, using a GB case study with one million domestic heat pump installations per year, that the sensitivity of electrical system demand to temperature (and subsequently sensitivities to cold/warm winter seasons) could increase by 50\% following four years of heat demand electrification. A central estimate of 1.75~kW additional peak demand per heat pump is estimated, with variability across three published heat demand profiles leading to a range of more than 14~GW in the most extreme cases. It is shown that the legacy approach of scaling historic demand, as compared to the explicit modelling of heat, could lead to over-procurement of 0.79~GW due to bias in estimates of additional capacity to secure. Failure to address this issue could lead to £100m overspend on capacity over ten years.
\end{abstract}


\end{frontmatter}

\section{Introduction}

Capacity markets have become a common framework for providing security of supply in energy systems without problems of oversupply of costly peaking capacity or the high economic and political costs of load shedding. Peak electricity demands are expected to grow significantly in regions with high levels of gas-fueled space heating, as heat is moved onto electrical systems to meet decarbonisation targets. These heat demands have much stronger sensitivities to meteorological and seasonal factors than historic electrical demand \cite{eggimann2019high,eggimann2020weather,staffell2018increasing,wilson2013historical}. In some countries this issue is compounded by the replacement of aging conventional thermal plant by renewable generation which is highly sensitive to meteorological conditions \cite{bloomfield2016quantifying,staffell2018increasing}. The study of system adequacy uses probabilistic methods, and demand uncertainty can both bias and increase variability in the estimates of capacity to secure; this bias should therefore be identified and, wherever possible, minimized. In fact, it is the \textit{range} of calculated capacity to secure values across scenarios that sets the procurement target in some markets \cite{ngeso2017ecr}. Given the large sums allocated in the capacity market (£700 million in the most recent GB capacity auction), even modest reductions in uncertainty could yield significant dividends in terms of social welfare. 

Despite this, the consideration of uncertainty in underlying changes in demand on system adequacy is seldom considered in detail. The system adequacy literature of the past decade has primarily focused on the determination of capacity value of non-dispatchable plant (e.g., renewables, demand-side response, energy storage) \cite{soder2020review,cole2020considerations,dent2016capacity,edwards2017assessing,nolting2020can,khan2018demand,lynch2019impacts,zachary2019integration}. These approaches assume a known, well-defined distribution of demand, with approaches typically scaling historic demand curves to meet projected forecast peak demand \cite{ngeso2017ecr}, neglecting changes in the distribution of demand duration curves that may occur due to clean energy transitions. To address this, a recent review of GB capacity market by independent academic experts made the formal recommendation that this issue be explored, noting that \textit{`The factors affecting the evolution of peak behaviour should be analysed ... from the broad perspectives of current and future technical, society and regulatory evolutions'} \cite{pte2020report}. This point is particularly pertinent given recent annual heat pump installation targets of 600,000 and one million per year (by the close of this decade) which have been proposed by the UK government and UK's Climate Change Committee, respectively; this will increase demand-weather sensitivity dramatically \cite{bossmann2015shape}.

Although there are a range of capacity market designs and time frames, it is generally the case that the most important timescale for capacity markets is typically between $\tau-1$ delivery for the following peak season (e.g., 18 months ahead of time for a summer auction to be delivered in the following winter) through to the $\tau-4$ delivery for the peak season four years ahead \cite{soder2020review}. This relatively short timescale with respect to energy transitions \cite{gross2019path,hall2016review,scamman2020heat} means that the practicalities of capacity procurement via markets (or other means) are not usually considered by papers that study the decarbonisation of heat; works usually focus on the increase in seasonality or growth of system peak over several decades. For example, the impacts of increasing heat pump demand are considered for the year 2050 in \cite{eggimann2019high}, estimating that 50\% penetration of heat pumps would result in increases in demand of more than 20~GW. Similarly, the works \cite{clegg2019integrated,clegg2019integrated2} develop a spatial heat model which is combined with network models for the GB gas and electric system, modelling prices and heat demand to 2050. It is estimated that peak demands could grow by close to 10~GW between 2020 and 2025. Other works consider the impacts of heat without explicitly considering the timescale of changes that could occur, such as \cite{wilson2013historical}. It demonstrates that, if 30\% of gas demand is shifted to highly efficient heat pump load, then electrical peak demands would increase by 25\%. The authors of \cite{cooper2016detailed} find that an 80\% heat pump demand scenario increases GB demand by as much as 54~GW in an unmitigated scenario, dropping to 16~GW under more favourable conditions. In \cite{eyre2015uncertainties}, the authors study four heat transition scenarios, noting that the pathways with widespread electrification of heat (via heat pumps) leads to particularly strenuous impacts on electricity systems. All of the aforementioned works point to electrification of heat causing systemic change, with significantly increased demand-weather sensitivity as well as base demand. Unfortunately, the disjuncture between the timescales of the capacity market and these works on long-term energy transitions results in challenges interpreting possible impacts on capacity procurement.

There are some works that do consider changes in demand up to five years ahead, although the implications of these works are not studied in the context of impacts on capacity markets. In \cite{staffell2018increasing}, the authors consider a five-year forecast of demand net of renewables, and (with the scenarios considered) estimate that net demand could drop by up to 5\%; the coefficient of performance (COP) of heat pumps is shown to be a key parameter that will determine the future impacts of heat electrification. The authors of \cite{eggimann2020weather} consider a similar problem, estimating changes in demand every five years until 2050. Long-term peak demand forecasting focuses on the problem of estimating the demand peak, using data-driven methods \cite{hyndman2009density,li2019use,khuntia2016forecasting}. These approaches do not explicitly consider the system margin or the subsequent capacity required to meet this demand; additionally, it is also usually implicit in data-driven models that demand growth can be extrapolated, with the main uncertainties typically in terms of economic growth (and so they are not appropriate if there is systemic change). In \cite{deakin2020calculations}, it is shown that the seasonality of electrified gas (heat) demand leads to 60\% faster peak demand growth than the equivalent electrical energy demand. However, across reviewed works, it has been identified that there is a gap in the analysis of impacts of decarbonisation of heat on capacity adequacy over the time frame of capacity markets, despite the key role that electrical heating is expected to play in achieving net zero.

This paper studies the impact of increasing the weather-dependency of electricity demand via electrification of space-heating, and how this will affect capacity markets through changes in capacity to secure. The paper develops a novel system adequacy model for this purpose which combines hourly demand and renewable generation models with an explicit space heating profile derived from daily gas demand and heat pump profiles. This enables the model to capture changes that heat pumps could have on demand profiles as well as on meteorological sensitivities. With this model, the impact of one million annual heat pump installations on a GB case study are considered comprehensively, with three specific aspects considered. 
\begin{itemize}
\item \textit{Changes in demand-weather sensitivity.} The evolution of demand-weather sensitivities are studied using Lasso-Regularized linear regression. Weather variables known to correlate with either gas or electricity demand are studied comprehensively, with the covariates subsequently derived from climate reanalysis data.
\item \textit{Bias arising from the legacy Load Duration Curve approach.} Models accounting for heat demand explicitly are compared against the legacy approach whereby heat demand growth is implied by the scaling of historic demand. Possible biases introduced by the legacy implicit approach are quantified.
\item \textit{Estimating scenario variability in Additional Capacity to Secure.} Scenario analysis is used to capture possible changes in variability of capacity to secure, with meteorological sensitivities and heat demand profile uncertainties considered.
\end{itemize}
By considering a detailed system adequacy model with these objectives in mind, we study not only the effects of heat pumps on demand, but also how they influence the capacity required to meet security of supply standards.

The contributions of the work are summarised as follows.
\begin{enumerate}
\item A demand model is proposed that explicitly accounts for increased electrical space heating demand at a national level, and is suitable for consideration within system adequacy studies. Space heating demand is estimated by assimilation of historical gas demand data with heat pump usage profiles.
\item The demand model is considered alongside 30 years of historic climate reanalysis data to create a demand hindcast covering winters from 1990 to 2020. The model uses Lasso-Regularized regression to avoid overfitting and exclude uninformative covariates. Net demand across each winter can then be hindcast using coincident renewable generation.
\item The model of net demand is combined with models of conventional and renewable generation to quantify security of supply in terms of loss of load expectation and subsequently capacity to secure for a GB case study. Scenario analysis across heat pump profiles and coefficients of performance show variability in capacity to secure greater than all scenarios presently considered in the most recent GB capacity market auction.
\item It is demonstrated for the first time that significant bias in capacity to secure could be introduced if models fail to capture changes in the underlying end-uses of electrical demand. This is achieved by comparing the explicit space heating model with conventional approaches that ignore changes in time- and weather-based dependencies of electrified heating demand.
\end{enumerate}

This paper has the following structure. Firstly, the novel Explicit heating model is introduced in Section \ref{s:adequacy}, to illustrate how heating demand can be accounted for in time-collapsed adequacy models in a natural way. In Section \ref{s:demand_weather}, we outline the Lasso-based linear regression approach, used to consider how heating demand could change the nature of future demand curves. The specific details of the GB system model are outlined in Section \ref{s:supply}, to introduce the key characteristics of the subsequent detailed case study. In Section \ref{s:results}, the full case study is used to study the key impacts of increased sensitivity, bias in capacity to secure estimates, and increased variability in capacity to secure. Salient conclusions on the modelling approach are drawn in Section \ref{s:conclusions}.

\section{Time-Collapsed System Adequacy Modelling}\label{s:adequacy}

A energy system is \textit{adequate} at a given time instant if there is sufficient generation to meet demand. Power systems are designed so that if all generation is available then there will always be a positive margin; however, due to unforced outages at generators and varying meteorological conditions (in systems with significant amounts of varying renewables) there is a non-zero probability of a shortfall occurring.

\begin{figure}\centering
\includegraphics[width=0.95\textwidth]{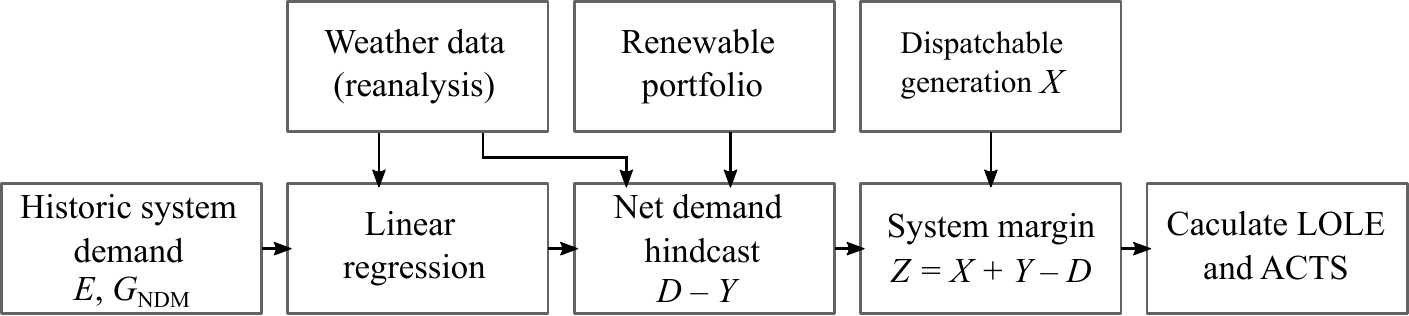}
\caption{High-level modelling approach. The net demand $D - Y$ is calculated for a given year by combining a linear model of historic system electrical and Non Daily Metered (NDM) gas demand $E,\,G_{\mathrm{NDM}}$ with coincident renewable generation $Y$. This is combined with dispatchable generation $X$, from which Loss of Load Expectation (LOLE) and Additional Capacity to Secure (ACTS) can be calculated. The NDM gas demand is used as a proxy for space heating demand.}\label{f:ae_structure}
\end{figure}

In this Section, we consider the structure of the full adequacy modelling approach, as summarized in Figure \ref{f:ae_structure} (with the exception of the linear regression approach which is considered in more detail in Section \ref{s:demand_weather}).

\subsection{Time-collapsed Adequacy Model}
A \textit{time-collapsed} (or `snapshot') adequacy model is designed to model the distribution of the system margin at a randomly selected time instant during the peak season \cite{ofgem2013electricity,zachary2019integration}. We propose the use of an \textit{hourly} time-collapsed model, where each hour of the day $t$ is modelled separately. The key advantage of this approach is that the impacts of a range of demand profiles on the whole day can be considered--this is important as some heat demand profiles peak in the morning (e.g., \cite{love2017addition}). Using this approach, the system margin $Z_{t}$ for a given hour $t$ is given by the linear sum
\begin{equation}\label{e:margin}
Z_{t} = X_{t} + Y_{t} - D_{t}\,,
\end{equation}
where $X_{t}$ represents dispatchable generation, $Y_{t}$ represents renewable generation, and $D_{t}$ represents total system demand. (Each of the variables of \eqref{e:margin} are random variables.) Dispatchable generation $X_{t}$ consists of conventional thermal plant and interconnectors, whilst renewable generation $Y_{t}$ is modelled as a combination of onshore wind, offshore wind, and solar generation. The `overall' margin $Z$ combines the profile for the whole day, and as such can be written
\begin{equation}\label{e:margin_calc}
Z = \sum_{t=0}^{23} I_{t}Z_{t}\,,
\end{equation}
where $I_{t}$ is a binary variable with a value of unity if it is hour $t$, and is zero otherwise.

It is worthwhile stressing the time dependencies in \eqref{e:margin}. Dispatchable generation $X_{t}$ is considered to be equally likely to be available during the whole peak demand period, and so this random variable is independent of demand and renewables \cite{wilson2018use}. On the other hand, both renewable generation $Y_{t}$ and demand $D_{t}$ are dependent on weather, and so the distribution of net demand $D_{t} - Y_{t}$ must be found by considering coincident times (i.e., these are both assumed dependent on the weather of a given time). Once the net demand has been found, the distribution of the system margin $Z_{t}$ for each hour $t$ can be found by convolution of the probability distribution functions of the net demand $D_{t} - Y_{t}$ and conventional generation $X_{t}$.

\subsubsection{Evaluating Security of Supply}
The system margin $Z$ can be used to define a range of risk metrics to understand the likelihood and severity of shortfalls. The likelihood is considered using the Loss of Load Expectation (LOLE), having units of hrs/yr, and is given by
\begin{equation}\label{e:lole}
\mathrm{LOLE} = \mathbb{E} \left ( \sum_{i=0}^{n-1} \mathrm{Pr}(Z < 0) \right )\,,
\end{equation}
where $n$ is the number of periods in year and $\mathbb{E}$ denotes the expectation operator. The LOLE metric has the advantage of being the target security standard of many European systems, whilst also being closely linked to the Loss of Load Probability (LOLP), which is used as an operational indicator of scarcity by transmission system operators. 

The LOLE is subsequently used to determine the Additional Capacity to Secure (ACTS). For a given security standard of $T_{\mathrm{LOLE}}$ hours per year, the ACTS is the (perfectly reliable) generation required to bring the LOLE to that security standard,
\begin{equation}\label{e:acts}
Z_{t}' = X_{t} + Y_{t} - D_{t} + \mathrm{ACTS} \quad \mathrm{s.t.} \quad \mathrm{LOLE}(Z') = T_{\mathrm{LOLE}} \,.
\end{equation}
where we use $\mathrm{LOLE}(Z)$ to denote the calculation of LOLE from \eqref{e:lole} using system margin $Z'$ (as calculated as in \eqref{e:margin_calc}).

For example, suppose that the GB security standard is 3 hours LOLE per year, but the generation already committed for the $\tau-4$ delivery year (the 24/25 winter) results in an LOLE of 10 hours per year, such that additional capacity is required. If, say, 1500 MW of perfectly reliable generation could bring the LOLE to 3 hours per year, then the ACTS would be 1500 MW. Note however that real generators are not perfectly reliable, and so a de-rating factor must be applied (i.e., more than 1500 MW of real generating capacity would need to be procured).

\subsubsection{Definition of Peak Demand}
It is also useful to define the peak demand for a season of $N_{\mathrm{Wtr.}}$ winter days so that the peak of different demand distributions can be compared. This work uses the method that is used to define Average Cold Spell peak demand \cite{ng2021acs}. This approach resamples winter demands many times to to determine the distribution of peak demands empirically; the median value of these demand peaks is then selected as the Peak Demand. This method can be denoted for the time-collapsed model of this work as
\begin{equation}\label{e:def_peak}
\mathrm{Peak\,Demand} = \mathrm{Median}\left( \max \{D_{0},\,D_{1},\,\cdots,\,D_{24N_{\mathrm{Wtr.}} - 1}\} \right) \,,
\end{equation}
where the $i$th random variable over which the $\max\{\}$ function is taken, $D_{i}$, has the corresponding demand model of that hour's margin (e.g., the 6am model is used for $D_{6},\,D_{30},\,$ and so forth).

\subsubsection{Impacts of ACTS on Capacity Markets}

The design of an effective capacity market is a challenge from both a practical and theoretical point of view \cite{cramton2005capacity,cramton2017electricity}. The GB capacity market is regarded as a well-designed, modern market \cite{newbery2016missing}, although technical details continue to develop, with annual recommendations from an independent Panel of Technical Experts \cite{pte2020report}. 

A brief overview of the design of this market is given in \cite{hawker2016capacity}. The Target Capacity to Secure is calculated using a range of supply- and demand-side sensitivities considered around the base case, as well as system-based sensitivities based on National Grid's Future Energy (FE) Scenarios. In total, between 20 and 30 sensitivities are typically considered. Based on these scenarios, the Least Worst Regret (LWR) methodology is used to estimate the Target Capacity to Secure from all scenarios, looking to identify the generating capacity that will have the smallest regret based on projected costs associated with oversupply (based on the net Cost of New Entry) and the costs of shortfall. The latter costs are calculated by the Expected Energy Unserved multiplied by a monetary estimate of the Value of Lost Load. Cost curves for each scenario are combined, and the aggregate least worst-regret option identified.

Explicitly calculating the LWR Target Capacity to Secure is beyond the scope of the current work: suffice to say, the Target Capacity to Secure is almost entirely dependent on only the largest and smallest estimates of the ACTS \cite{ngeso2017ecr}. As such, not only is it important that calculations of ACTS have low bias, but also that the \textit{variability} in forecasts for supply and demand are correctly captured. The Range of Capacity to Secure (RoCS) is therefore considered to evaluate the total variability across all scenarios, and is defined as
\begin{equation}\label{e:rocs}
\mathrm{RoCS} = \max_{\Sc} \{\mathrm{ACTS}(\Sc ) \} - \min_{\Sc} \{\mathrm{ACTS}(\Sc ) \}\,,
\end{equation}
where $\mathrm{ACTS}(\Sc)$ denotes the calculation of capacity to secure ACTS using scenario $\Sc$. For a more detailed critical discussion on the LWR methodology see \cite[Apdx. 7]{ngeso2017ecr}.

\subsection{Explicit System Model, using Non Daily Metered Gas Demand $G_{\mathrm{NDM}}$ as a Proxy Heat Variable}\label{ss:explicit}

We first consider a system model with electrical demand $D_{t}$ which is explicitly decomposed into underlying electrical demand $E_{t}$ and electrified space heating demand $H_{t}$, as
\begin{equation}\label{e:D_diaggr}
D_{t} = E_{t} + H_{t}\,.
\end{equation}
Given this disaggregtion, we refer to the model \eqref{e:D_diaggr} (combined with \eqref{e:margin}) as the \textit{Explicit} system model, as heating demand is accounted for explicitly within the calculations of system adequacy.

For the purposes of this work, the Explicit model \eqref{e:D_diaggr} will be considered the ground truth (for a given heat demand model $H_{t}$), to which alternative approaches will be considered. This is because space heating demand $H_{t}$ has been observed to have a very different demand profile to that of underlying electrical demand $E_{t}$, irrespective of whether the means of fulfilling that demand is by gas boilers or electric heat pumps \cite{love2017addition,wilson2013historical}. The approach therefore has advantages of being more closely linked to the systemic changes driven by the electrification of heat, although a method of estimating the electrical heating demand $H_{t}$ is required.

\subsubsection{Estimating Electric Space Heating Demand}
To model heating demand $H_{t}$ for the Explicit system model, we propose that suitably scaled Non Daily Metered (NDM) gas demand $G_{\mathrm{NDM}}$ is used as a proxy variable, in a similar vein to \cite{deakin2020calculations,clegg2016assessment}. NDM gas demand is largely comprised of water and space heating demand (cooking using gas is less than 3\% of total domestic consumption) \cite{eurostat2018questionnaire}, and the customer composition is largely residential and flats/commercial properties. It does not include large industrial customers (such as gas-fired power stations), who are instead billed as part of the Daily Metered class \cite{ng2016gas}.

Following other works, it is assumed that the daily electrified heating demand follows some electrified heating profile $h$, and that water heating demand $G_{\mathrm{HW}}$ is approximately constant throughout the year \cite{ruhnau2019time}. As such, the (space) heating demand $H$ is calculated as an affine function of the daily NDM gas demand $G_{\mathrm{NDM}}$ as
\begin{equation}\label{e:gas2heat}
H_{t} = \dfrac{h_{t} n_{H} f_{\mathrm{Dom}}}{k_{\mathrm{COP}}} \left ( G_{\mathrm{NDM}} - G_{\mathrm{HW}}\right ) \,,
\end{equation}
where $G_{\mathrm{HW}}$ is the (constant) daily hot water demand for gas, $f_{\mathrm{Dom}}$ is the fraction of gas demand meeting domestic demands, $k_{\mathrm{COP}}$ a system-wide coefficient of performance, and $n_{H}$ is the number of customers with electrified heating demand.

\subsection{Implicit System Model, using Load Duration Curves}\label{ss:implicit}

The standard approach for considering the evolution of system demand is via the use of load duration curves. With this approach, the estimated peak demand for a given year $k_{\mathrm{Peak}}$ is forecast by some means. Once this has been determined, the distribution of the total demand $D$ is calculated by linearly scaling the electrical demand $E$ as
\begin{equation}\label{e:ldc_peak}
D = k_{\mathrm{Peak}} E\,.
\end{equation}
The model \eqref{e:ldc_peak}, combined with the system margin \eqref{e:margin} we refer to as the \textit{Implicit} system model, as changes in heat demand are implied by the coefficient $k_{\mathrm{Peak}}$. This approach is therefore used, for example, in the methodology for modelling demands in the GB Capacity Market \cite{ngeso2019ecr}. For the purposes of creating Implicit models that are equivalent to Explicit demands in a meaningful way, $k_{\mathrm{Peak}}$ is chosen so that the Peak Demands \eqref{e:def_peak} are equal.

The clear advantage of this approach is its conciseness: once suitable load duration curves $E$ have been identified, only the peak demand coefficient $k_{\mathrm{Peak}}$ for a given year needs to be determined. On the other hand, it must be assumed that the load duration curve describing the electrical demand $E$ will not change significantly. As mentioned in Section \ref{ss:explicit}, there are good reasons to think that heat demand $H_{t}$ has a different distribution to electrical demand $E_{t}$; however, if changes to electrified heat demand are small, then this Implicit model might be preferable. 

To study explicitly the differences between the models, we consider the Bias in the estimates of ACTS for a given scenario $\mathcal{S}$ to be given by
\begin{equation}\label{e:bias}
\mathrm{Bias}( \mathcal{S} ) = \mathrm{ACTS}_{\mathrm{Im.}}( \mathcal{S} ) - \mathrm{ACTS}_{\mathrm{Ex.}}( \mathcal{S} ) \,,
\end{equation}
where the subscripted $\mathrm{ACTS}_{\mathrm{Ex.}},\,\mathrm{ACTS}_{\mathrm{Im.}}$ are the calculations of the ACTS using Explicit model \eqref{e:D_diaggr} and Implicit model \eqref{e:ldc_peak}, respectively. In this way, the effect of changes in demand profiles on the ACTS can be taken into account for models which otherwise are identical according to their Peak Demand \eqref{e:def_peak}.

\section{Weather-Dependent Energy System Modelling}\label{s:demand_weather}

A variety of statistical inference procedures, aimed at understanding the effects of exogenous factors (such as weather) on energy demand, have been developed by both academia and industry. To ensure that all possible effects of increased space heating are captured, we consider the statistical inference methods developed by both the gas system operator, National Grid Gas, and the electricity system operator National Grid ESO (NGESO).

NGESO estimates the sensitivity of electricity demand to weather using the Average Cold Spell methodology \cite{ng2021acs}. This calculates the sensitivity of unrestricted system demand to weather variables. (Unrestricted system demand is defined as the sum of the transmission system demand with and demand-side response, embedded generation and interconnector exports all accounted for.) From this, the `underlying', non-weather sensitive demand can be estimated. The exact weather variables used are not specified, unfortunately, but by far the most common weather variable studied in academic literature is temperature (alongside temporal variables such as day of week) \cite{hyndman2009density,de2015seasonal,hong2016probabilistic,hilbers2019importance,eggimann2019high,eggimann2020weather,wilson2018use,bloomfield2020characterizing,bloomfield2016quantifying}.

National Grid Gas uses the Composite Weather Variable \cite{xoserve2014autumn} to quantify the impacts of weather on demand. Unlike the Average Cold Spell methodology, however, many of the variables used in the Composite Weather Variable are public, and are described in \cite{desc2019review}. The variables include temperature, wind chill, solar irradiance, and in future could include the effects of precipitation. Combining the approaches from gas and electrical domains, a total of ten weather-based covariates are considered, as well as a range of temporal variables (see Table \ref{t:weather_variables}).

\begin{table}
\centering
\begin{tabular}{ll}
\toprule
Variable & Description\\
\midrule
$W_{\mathrm{on}},\,W_{\mathrm{off}}$ & Hourly onshore/offshore wind capacity factors \\
$S,\,\bar{S}$ & Solar PV capacity factor \\
$T,\,\bar{T}$ & Population-weighted temperature \\
$W_{\mathrm{Chill}},\,\bar{W}_{\mathrm{Chill}}$ & Population-weighted wind chill \\
$T_{\mathrm{Cold}},\,\bar{T}_{\mathrm{Cold}}$ & Population-weighted cold-spell uptick\\
$t_{\mathrm{Mon}}\,-\,t_{\mathrm{Sun}}$ & Binary variable for weekdays (each of Monday to Sunday)\\
$t_{\mathrm{Prd,\, C}i},\,t_{\mathrm{Prd,\, S}i}$ & $i$th harmonic time component (C/S as cosine/sin terms)\\
$t_{\mathrm{Sunset}}$ & Sunset time\\
$t_{\mathrm{Lin}}$ & Linear time variable\\
$\hat{E}_{\mathrm{Out}}$ & Out-turn peak electrical demand\\
$\hat{G}_{\mathrm{Out}}$ & Out-turn mean winter gas demand\\
\bottomrule
\end{tabular}

\caption{Summary of weather variables and covariates used for energy system simulation, each considered at an hourly resolution. Variables with a bar $\bar{(\cdot)}$ average the variables for the previous 24 hours.}\label{t:weather_variables}
\end{table}

\subsection{Converting Meteorological Reanalysis to Weather-Based Covariates}

Accounting for the dependencies of energy systems on weather requires an approach to convert historic weather measurements into appropriate covariates. \textit{Meteorological reanalyses} are becoming increasingly popular for modelling of weather within the context of energy systems. These datasets are a gridded reconstruction of past weather observations, created by combining historic observations with a high-fidelity numerical model of the earth system, providing a high quality, comprehensive record of how weather and climate have changed over multiple decades. In the context of energy systems, the high spatial resolution allows for the climate of a region or country to be captured, as well as being freely available for researchers \cite{staffell2016using,hilbers2019importance,bloomfield2018changing,cannon2015using}. In this work, we develop the methods described in \cite{bloomfield2020characterizing,bloomfield2020merra} for deriving weather-based covariates from the MERRA-2 reanalysis data \cite{gelaro2017modern}.

\begin{figure}\centering
\subfloat[Offshore Wind]{\includegraphics[width=0.3\textwidth]{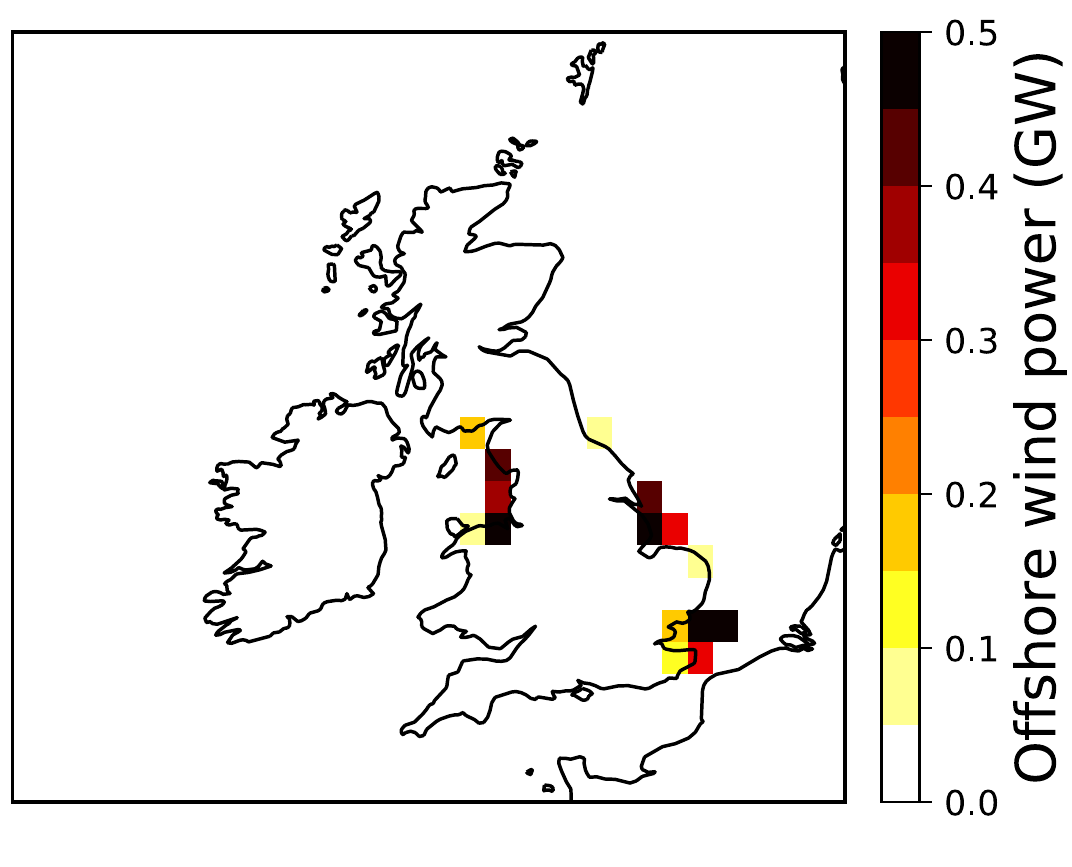}\label{f:GB_ofs_wind_updated}}
~
\subfloat[Onshore Wind]{\includegraphics[width=0.3\textwidth]{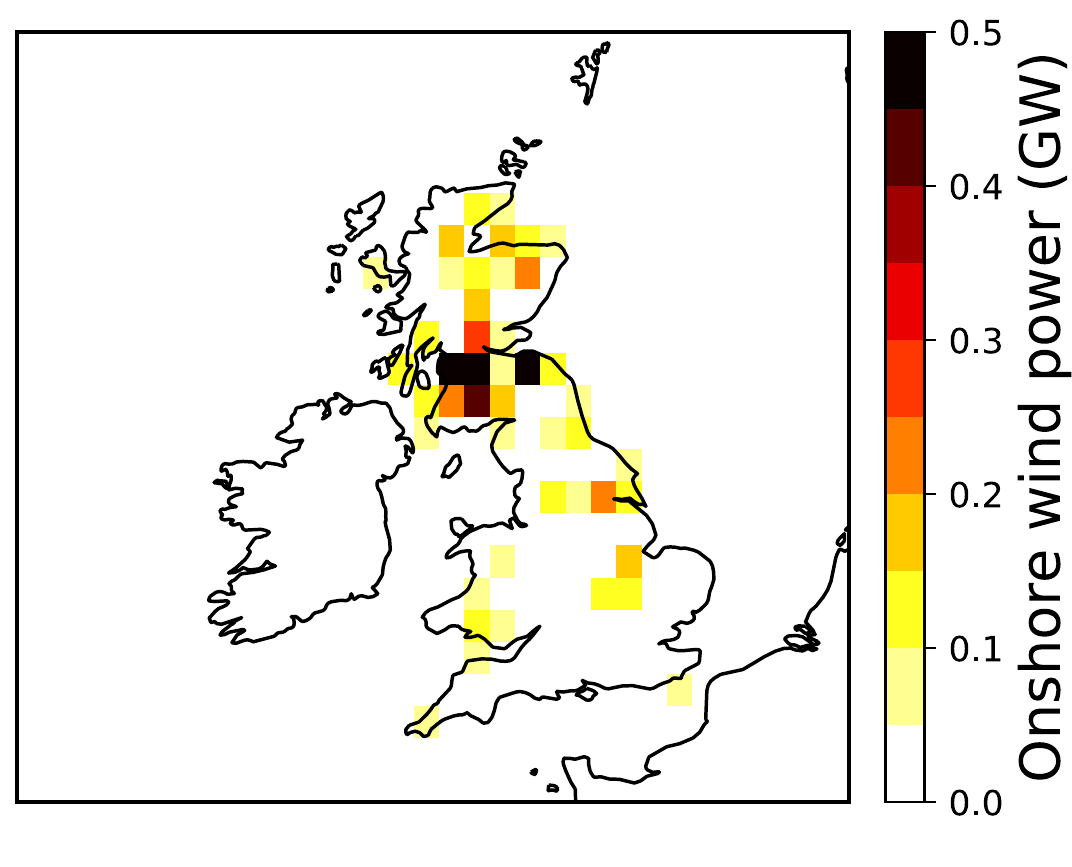}\label{f:GB_ons_wind_updated}}
~
\subfloat[Population]{\includegraphics[width=0.3\textwidth]{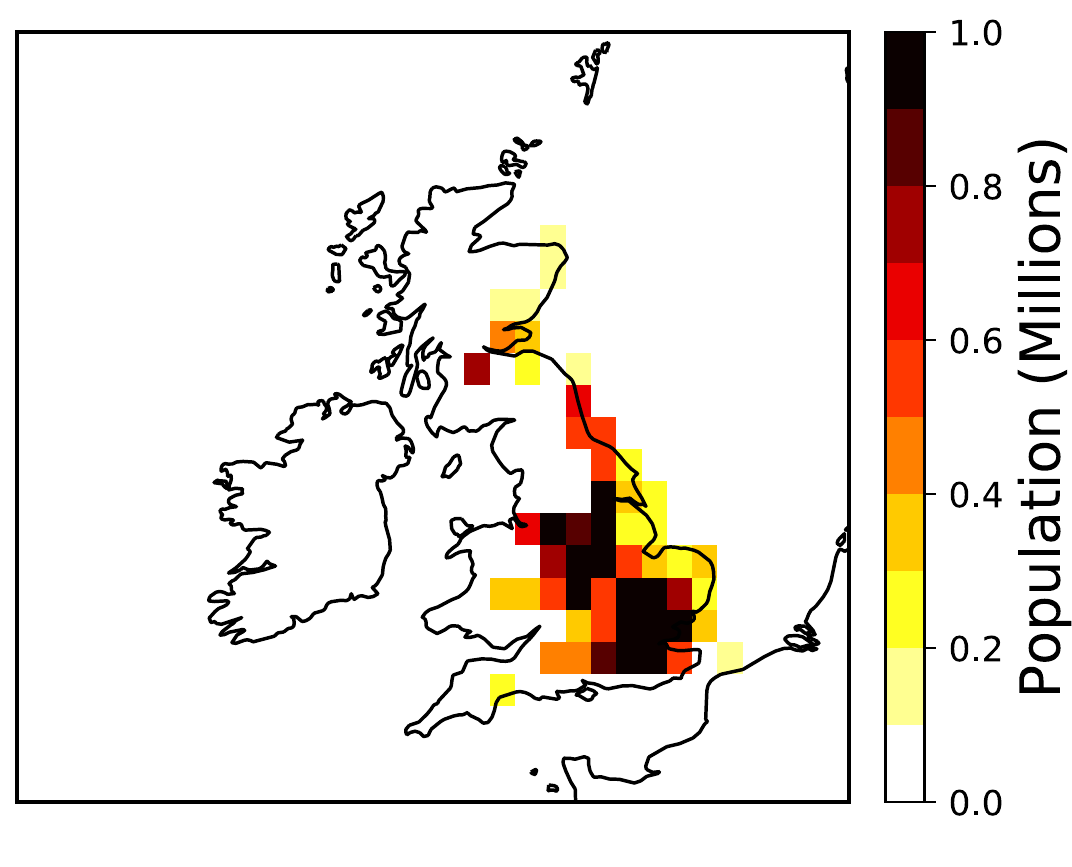}\label{f:population_weights}}
\caption{The onshore and offshore wind locations (a, b) and the population weights (c) for Great Britain, used for converting raw reanalysis weather data into wind capacity factors and weather covariates.}
\label{f:figsAandB}
\end{figure}

In total, there are six covariates based on meteorological conditions (and a further four averaged variables), as in Table \ref{t:weather_variables}. These parameters are derived from the raw reanalysis data as follows:
\begin{itemize}
\item Hourly onshore and offshore wind capacity factors $W_{\mathrm{On}},\,W_{\mathrm{Off}}$ are constructed by modification of the method from \cite{bloomfield2020characterizing}, with the main steps summarised here. The MERRA-2 reanalysis near-surface wind speeds are extrapolated using a power-law to 58.9~m and 85.5~m, based on the capacity-weighted mean onshore and offshore turbine hub heights respectively (from \cite{windpower2020}). Onshore and offshore turbine power curves from \cite{ngeso2019derating} are obtained using \cite{Rohatgi2020}, with aggregated GB onshore and offshore wind capacity factors then created by considering wind farm locations from \cite{windpower2020} for the year 2017 (Figures \ref{f:GB_ofs_wind_updated}, \ref{f:GB_ons_wind_updated}). As in \cite{bloomfield2020characterizing}, this results in good accuracy compared to out-turn data \cite{entsoe2020transparency}, with Coefficient of Determination $R^{2}$ of 0.95, 0.90 and RMS error of 7.9\%, 7.5\% for onshore and offshore wind capacity factors respectively, when compared against 2018 daily forecast capacity factors.
\item Hourly solar capacity factors $S$ are modelled using a combination of surface temperature and incoming surface irradiation, as described in \cite{bloomfield2020characterizing}.
\item Hourly temperature $T$ and cold-weather uptick variables $T_{\mathrm{Cold}}$ are calculated based on population-weighted 2 meter temperatures. (The latter is used as a proxy for the Cold Weather upturn of the Composite Weather Variable, and is intended to model increased demand at cold temperatures.) The cold-weather uptick is calculated as
\begin{equation}\label{e:t_chill}
T_{\mathrm{Cold}} = \max \{T_{0} - T,0\}\,,
\end{equation}
with a cold spell cut-off temperature $T_{0}$ of 3\degC, based on \cite[pp. 6]{desc2019seasonal}. The population weighting of GB is shown in Figure \ref{f:population_weights}.
\item Wind chill is calculated by multiplying the population-weighted 2-meter temperature $T$, with respect to a wind chill temperature $T_{\mathrm{WC}}$ against a similar population-weighted 2-meter wind speed $W_{\mathrm{Pop}}$, also with respect to a threshold $W_{\mathrm{WC}}$
\begin{equation}\label{e:wchill}
W_{\mathrm{Chill}} = \max \left \{T_{\mathrm{WC}} - T,0 \right \} \times \max \left \{W_{\mathrm{Pop}} - W_{\mathrm{WC}}, 0\right \}\,.
\end{equation}
The wind chill parameters $W_{\mathrm{WC}},\,T_{\mathrm{WC}}$ are chosen as $-1.5$ m/s and 16.5\degC, which are consistent with regional values reported in \cite[pp. 6]{desc2019seasonal}.
\end{itemize}

\subsection{Linear Regression with Lasso Regularization}\label{ss:linear_regression}

The goal of regression is to determine underlying sensitivity of a dependent output variable with respect to given input variables (covariates). Least-squares linear regression typically achieves this goal by minimizing the square of the residuals, with the Least Square sensitivities $\theta_{\mathrm{L.S.}}$ determined as
\begin{equation}\label{e:theta_ls}
\theta_{\mathrm{L.S.}} = \arg \min_{\theta} \|y - \theta^{T}x \|_{2} \,,
\end{equation}
for covariates $x$ and output $y$. Unfortunately, na\"{i}ve Least Squares \eqref{e:theta_ls} can lead to over-fitting as there is no penalty on the complexity of a model, risking returning a model with poor predictive performance due to spurious correlations \cite[Ch. 7.2]{hastie2009elements}. Indeed, with the relatively large number of covariates considered in this work, it was found that the models showed poor out-of-sample predictive capabilities compared to within-sample fitted data.

To overcome this issue, we use Lasso Regularization \cite[Ch. 3]{hastie2009elements}. In addition to mitigating against over-fitting, Lasso Regularization provides solutions $\theta_{\mathrm{Lasso}}$ that are \textit{sparse}. That is, coefficients corresponding to covariates which have little or no impact on the output are set to zero.

This is achieved by adding a regularization term $\alpha \|\theta \|_{1}$ to the Least Squares cost function \eqref{e:theta_ls}, with the Lasso estimate of the sensitivities $\theta_{\mathrm{Lasso}}$ determined as
\begin{equation}\label{e:theta_lasso_a}
\theta_{\mathrm{Lasso}}( \alpha ) = \arg \min_{\theta} \|y - \theta^{T}x \|_{2} + \alpha \|\theta \|_{1} \,.
\end{equation}
The regularization term penalizes large coefficients, having the effect of reducing the magnitude of individual entries in $\theta_{\mathrm{Lasso}}$, depending on the value of $\alpha$. For $\alpha \to 0$, the Lasso estimate tends to the Least Squares estimate \eqref{e:theta_ls}, and will therefore tend to over-fit; on the other hand, for sufficiently large $\alpha$, all values in the vector $\theta_{\mathrm{Lasso}}$ will be zero, under-fitting in most cases. Between these two extremes will be an `optimal' value of $\alpha$, $\alpha^{*}$, which will maximise the out-of-sample predictive performance (in this work Coefficient of Determination, $R^{2}$, is used as a scoring function). 

The optimal Lasso fit $\theta_{\mathrm{Lasso}}^{*}$ is fitted with this value of $\alpha$, i.e.,
\begin{equation}\label{e:theta_lasso}
\theta_{\mathrm{Lasso}}^{*} = \theta_{\mathrm{Lasso}}(\alpha^{*}) \,.
\end{equation}
Thus, a method is required to estimate the out-of-sample of predictive performance and subsequently determine $\alpha^{*}$.

\subsubsection{Determining the Lasso Parameter $\alpha^{*}$ and Computational Complexity}

To determine the out-of-sample predictive performance, and subsequently determine an optimal choice of $\alpha$, we use $k$-fold cross-validation \cite[Ch. 7.10]{hastie2009elements}. The approach can be briefly summarised as follows:
\begin{itemize}
\item $k$ cross-validation folds are created from the $k$-years of data, with each fold having one year of data for validation and $k-1$ years for training.
\item The sensitivity $\theta_{\mathrm{Lasso}}(\alpha)$ is determined for a range of values of $\alpha$ and for each of the $k$ cross-validation folds.
\item Estimates of the mean and standard error of the prediction score ($R^{2}$) are calculated for each value of $\alpha$ using the value of $R^{2}$ calculated for each of the $k$ cross-validation folds.
\item The value of $\alpha^{*}$ is chosen for which the mean prediction score is within one standard error of the maximum value of the prediction score $R^{2}$.
\item Finally, $\theta_{\mathrm{Lasso}}^{*}$ is calculated from \eqref{e:theta_lasso} (using data from all $k$ winters).
\end{itemize}

Although there is no closed-form solution to \eqref{e:theta_lasso}, the computational complexity of the Lasso is typically the same as ordinary Least Squares \cite[Ch. 3]{hastie2009elements}. The \texttt{scikit-learn} package \cite{pedregosa2011scikitLearn} is used for all regression calculations.

\section{GB Case Study System Modelling}\label{s:supply}

In this section we discuss how the energy data sources of Table \ref{t:dataSources_ae} are used to build and validate a generally representative model of the GB system. The GB system is chosen for study as it has had a capacity market functioning for several years, and because it has a very high fraction of its domestic space heating demand met currently met by natural gas \cite{eurostat2020energy}.

\begin{table}
\centering
\begin{tabular}{lll}
\toprule
Variable & Description & Reference/data \\
\midrule
$X$ & Dispatchable generation & \cite{ngeso2020fes,ng2013etsy,ofgem2014electricity,ofgem2020existing} \\
$E$ & Unrestricted electrical system demand & \cite{ngeso2020data,ngeso2020winter,national2019security,beis2020dukes}\\
$G_{\mathrm{NDM}}$ & NDM gas demand & \cite{ngg2020data} \\
$G_{\mathrm{HW}}$ & Hot water gas demand & \cite{eurostat2018questionnaire,ruhnau2019time} \\
$h$ & Electrified heat demand profiles & \cite{eyre2015uncertainties,love2017addition,watson2019decarbonising}\\
$k_{\mathrm{COP}}$ & System-wide heat pump COP & \cite{staffell2012review,eyre2015uncertainties} \\
$n_{H}$ & Fraction of houses converted to space heating & \cite{ccc2020sixth} \\
$f_{\mathrm{Dom}}$ & Domestic fraction of NDM gas demand & \cite{beis2019sub,ngg2020data} \\
\bottomrule
\end{tabular}

\caption{Summary of GB system case study data sources.}\label{t:dataSources_ae}
\end{table}

\subsection{Growth in Underlying Demand and Generation}

The peak demand season for both electrical and gas systems in Northwest Europe occurs during the winter months from November to March. Following prior works, we study of the 20 weeks of the year following the first Sunday of November, with the exception of the two weeks surrounding Christmas (these two weeks have low demand and so the likelihood of shortfall is negligible) \cite{zachary2014estimation,wilson2018use,sheehy2016impact}.

The underlying electrical demand $E$ is assumed to remain steady at 19/20 levels, following industry five-year forecasts \cite{ngeso2020fes}. To consider a rapid but credible increase in heat pumps in a system on top of this, as could be considered at some point on a pathway to net-zero, we consider a rate of one million domestic installations per year \cite{ccc2020sixth}.

Conventional generators are represented by a two-state model, using forced-outage rates from \cite[Table 1]{ofgem2014electricity}, with reported availabilities between 81\% and 97\% for these technologies. The forecast of total installed capacity of each class of generation technology is taken from the five-year forecast \cite{ngeso2020fes}; following previous works, these total values are then disaggregated into individual generating units based on unit sizes taken from National Grid's 2013 `Gone Green' scenario \cite{deakin2020calculations}. The distribution of interconnector flows are modelled with a uniform distribution. Specifically, imports are assumed to equally likely between high and low capacity factors of the individual interconnected countries reported by NGESO \cite{ngeso2020ecr}, with flows assumed independent.

\begin{figure}\centering
\subfloat[Demand, $D$]{\includegraphics[width=0.325\textwidth]{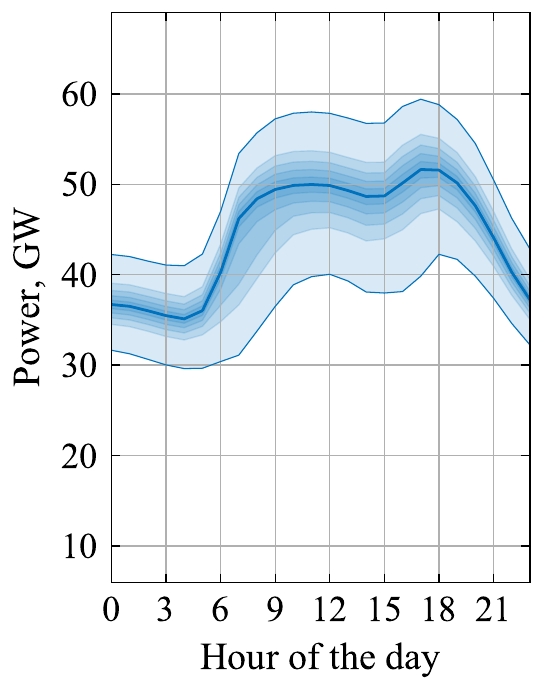}\label{f:plt_demand_deciles_yr0_L}}
~
\subfloat[Net Demand, $D - Y$]{\includegraphics[width=0.325\textwidth]{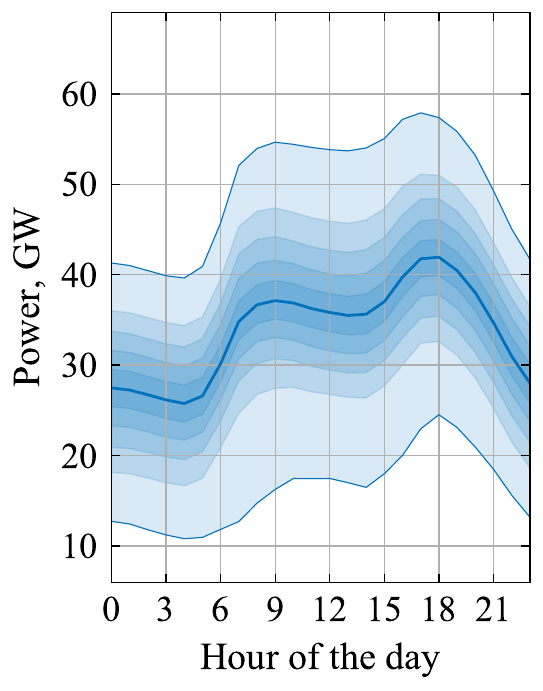}\label{f:plt_demand_deciles_yr0_N}}
~
\subfloat[Dispatchable Gen., $X$]{\includegraphics[width=0.325\textwidth]{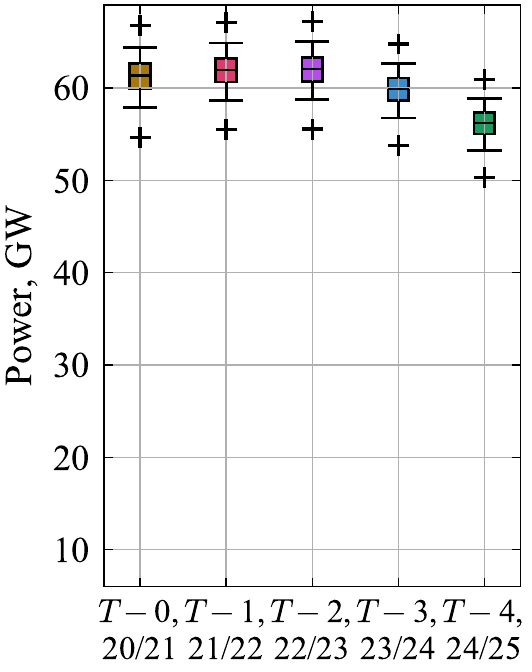}\label{f:pltGenBoxplots}}
\caption{The system has a peak demand (a) is close to 60~GW, with the variability of net demand (b) much greater due to variable renewable generation. (Plotted in (a), (b) are 10-90\% deciles, as well as the 1-in-10 year quantiles, based on a 30 year hindcast of demand and renewables.) As thermal plant retirements increase, the dispatchable generation $X$ (c) reduces. Boxplots show the 0.1, 5, 25, 50, 75, 95, and 99.9\% quantiles.}
\label{f:system_figs}
\end{figure}

With these models for interconnectors and conventional generators, the probability density of the dispatchable generation $X$ can be determined via convolution of all individual generators and interconnectors \cite{deakin2020calculations}. Boxplots of the distribution of the resulting random variable $X$ are shown in Figure \ref{f:pltGenBoxplots} for each delivery year. The median of the generation $X$ for 19/20 (discounting embedded generation) is 54.6~GW, compared to the previously procured capacity of 52.4~GW for 20/21 as reported in capacity market reports, and is therefore considered reasonably representative of the GB system.

\subsubsection{Estimating Historic System Demand}

The estimation of total demand is challenging due increasing levels of embedded generation \cite{beis2020dukes} and customer demand management (CDM--colloquially referred to as `triad avoidance', and results in up to 2.5~GW of demand-side response) \cite{ngeso2020winter}. Embedded generation represents a range of technologies, both dispatchable (such as small diesel generation) or renewable wind and solar generators. Additionally, NGESO keeps reserves to cover the loss-of-largest-infeed, which at present has a value of 1.32~GW \cite{national2019security} (this effectively increases demand by the same amount).

The unrestricted system demand $E$ is therefore determined as the sum of Transmission System Demand \cite{ngeso2020data}; estimated embedded wind and solar generation output \cite{ngeso2020data}; estimates of remaining embedded generation \cite{beis2020dukes} (assuming an availability of 90\%); and, estimates of customer demand management (provided by NGESO). The latter is assumed to run at 100\% from 5-6 pm and at 40\% at 4pm/7pm. Collating all data, peak demands (without weather correction) match NGESO estimates of weather-corrected peak to within 2~GW from 14/15 through to 19/20 winters with mean absolute error of 1.05 GW. The closeness of estimates gives confidence that the demand model $E$, like the generation model $X$, is also broadly representative of the GB system.

\subsection{Heat Pump Load Profiles and System-wide Coefficient of Performance}

\begin{figure}\centering
\includegraphics[width=0.84\textwidth]{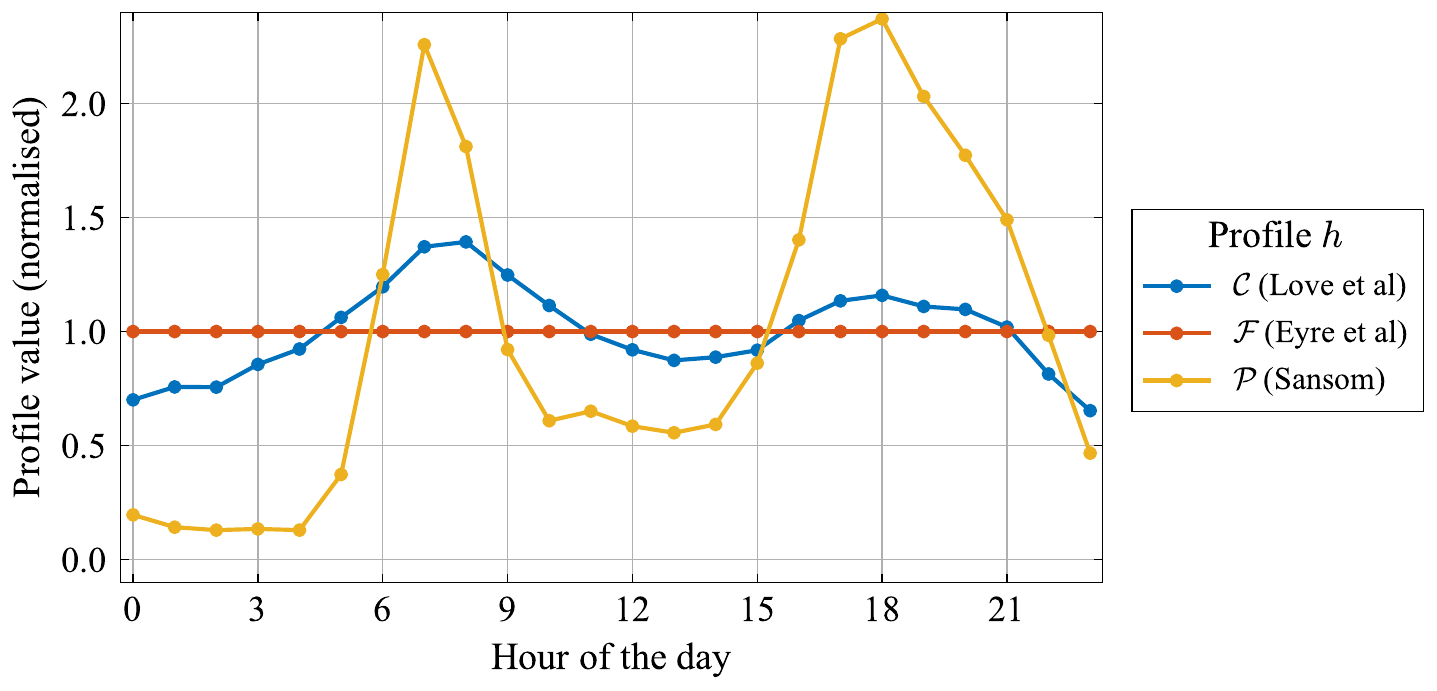}
\caption{The three heat pump profiles used for sensitivity analysis. The central profile $\mathcal{C}$ is from Love et al \cite{love2017addition}; the flat profile $\mathcal{F}$ is from Eyre at al \cite{eyre2015uncertainties}; and the peaking profile $\mathcal{P}$ of Sansom is from \cite{watson2019decarbonising}.}\label{f:plt_hp_variations}
\end{figure}

The heat pump load profile $h$ will have a large impact on results, as heat demand $H$ at each hour is linearly related to this profile \eqref{e:gas2heat}. Therefore, three heat pump profiles are considered as system-wide sensitivities, taken from literature using \cite{Rohatgi2020}, and then normalised and compared in Figure \ref{f:plt_hp_variations}. The first of these we define as our central profile $\mathcal{C}$, and is taken as the cold-weather weekday profile of Love et al \cite{love2017addition}, and is based on measured data from several hundred UK-based heat pumps. Secondly, we consider the flat profile $\mathcal{F}$, as considered by Eyre et al in \cite{eyre2015uncertainties} (such a profile has been reported as the de-facto standard for heat demand in \cite{eggimann2019high}). Finally, we compare this against the profile $\mathcal{P}$ of Sansom, as described in \cite{watson2019decarbonising}, from hereon referred to as the `peaking' profile. It is noted in \cite{watson2019decarbonising} that $\mathcal{P}$ has been very influential in policy, even though it has a peak higher than other estimates of half-hourly heat demand.

The system-wide coefficient of performance $k_{\mathrm{COP}}$ is also subject to considerable uncertainty. In \cite{eyre2015uncertainties}, the authors estimate the value of the COP for air-source heat pumps could increase from 2.0 up to 3.0 from 2010 through to 2050, similarly increasing from 2.5 to 4.0 for ground-source heat pumps, although the authors assume the COP at peak to be 0.8 times lower than this (due to colder temperatures during peak demands). Similarly, in \cite{staffell2012review}, the authors estimate seasonal performance factors (equivalent to the COP definition used in this work) of 1.5-2.1 for air-source heat pumps and 2.0-2.8 for ground-source heat pumps; again, during cold weather the performance of these systems will likely be lower than these values. To capture the range of values and possible improvements in building stock during any refitting, we therefore consider a COP range from 1.5 to 2.8, with a central estimate of 2.0.

The fraction of NDM gas demand used by domestic customers $f_{\mathrm{Dom}}$ is 79\% \cite{beis2019sub,ngg2020data}. It is assumed that hot water heating demand $G_{\mathrm{HW}}$ is evenly spread through the year \cite{ruhnau2019time} and that commercial and domestic properties have a similar use of hot water. Under these assumptions, the mean hourly gas demand for hot water is 9.9~GW throughout the year \cite{eurostat2018questionnaire}.

\section{Results}\label{s:results}

The aim of this work is to consider how electrification of heat could impact on capacity markets through changes in Additional Capacity to Secure. In Section \ref{ss:regression}, we first demonstrate the Lasso Regularization approach (as described in Section \ref{ss:linear_regression}), before considering how the meteorological sensitivity could evolve for $\tau-4$ delivery in the 24/25 winter. Based on this model, in Section \ref{ss:bias} we then consider how bias could be introduced in estimates of ACTS by Implicit modelling, even when the equivalent Explicit model has identical peak demand levels. Finally, in Section \ref{ss:variability}, the variability in the estimates of ACTS are considered across a range of scenarios.

\subsection{Net Demand Regression with Meteorological Variables}\label{ss:regression}

\begin{figure}\centering
\includegraphics[width=0.96\textwidth]{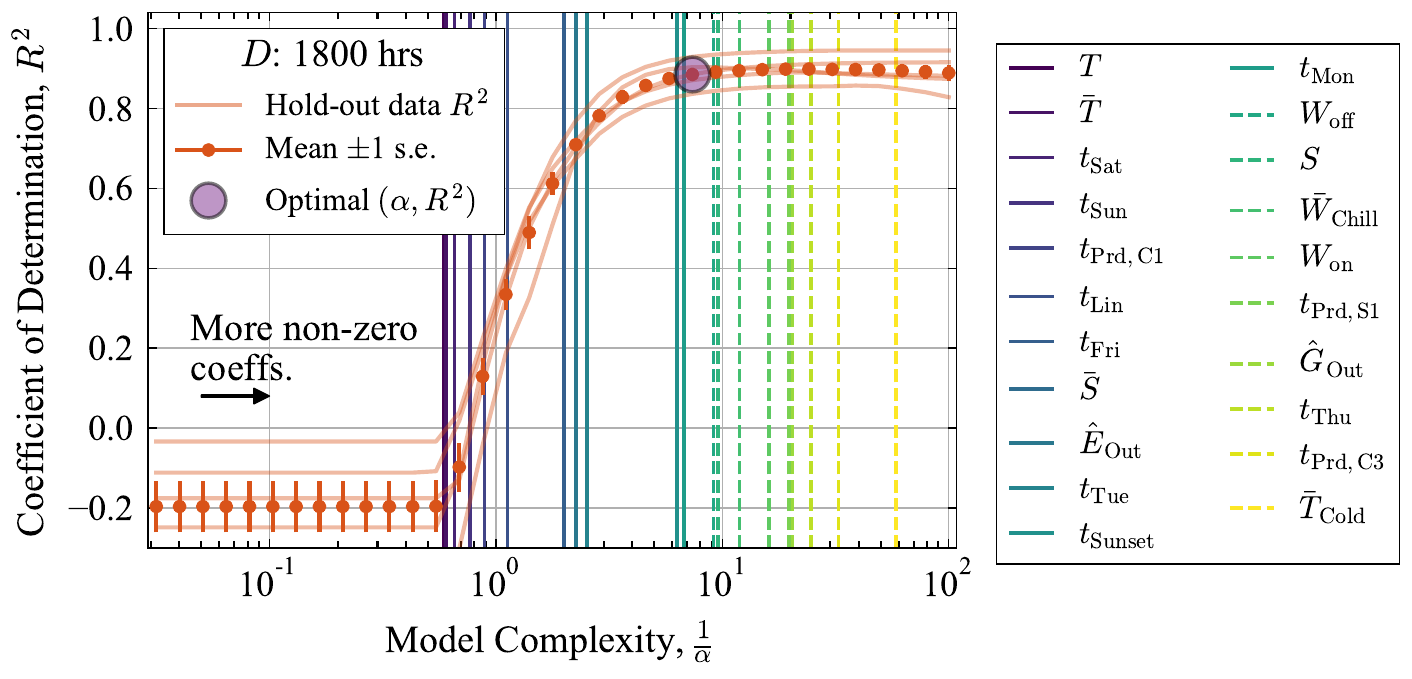}
\caption{Linear regression approach, using Lasso Regularization with 5-fold cross-validation. As the regularization weight decreases (so that $1/\alpha$ increases), the number of non-zero coefficients increases, as indicated by the vertical lines. The optimal value of $\alpha$, $\alpha^{*}$, is chosen at the point at which the mean coefficient of determination $R^{2}$ is within one standard error of the highest mean value of $R^{2}$.}\label{f:pltCompTest_vrbsD}
\end{figure}

We first illustrate the Lasso Regularization approach outlined in Section \ref{ss:linear_regression}, considering fitting the linear model for the system demand at 6pm for $\tau-0$ (the 20/21 winter, with no heat pump demand). Figure \ref{f:pltCompTest_vrbsD} outlines the approach: a model is fit for each of the five training datasets (each with one contiguous winter used as a hold-out validation set); from this, the mean and standard error of the Coefficient of Determination from the hold-out validation data are calculated. The optimal value $\alpha^{*}$ is selected as the value that is within one standard error of the maximum mean coefficient of determination from that hold-out scoring. Each vertical line indicates the value of $1/\alpha$ for which the coefficients of a given covariate becomes non-zero, from which it can be observed that there are many coefficients which are estimated to have a value of zero.

Using this approach for all hours, the Coefficient of Determination $R^{2}$ with the optimal Lasso parameter $\alpha^{*}$ varies between 0.73 and 0.94; it is above 0.87 for all hours periods between 6am and 9pm. The mean value of $R^{2}$ for each of these hours is above 0.8 using the out-of-sample validation data. This performs very favorably compared to the least squares model--although the mean Coefficient of Determination throughout the day was reduced by 5\%, the least squares modelling over-fit dramatically at most times, with all but a single time period having a Coefficient of Determination lower than $-8$ using the out-of-sample validation data. If the 24 models of each of the hours are combined, the overall coefficient of determination $R^{2}$ is 0.978 with sample standard deviation of all combined residuals of 1.02 GW  (1.69\%), indicating good performance.

\subsubsection{Meteorological Sensitivity from Explicit and Implicit Models}

To consider more clearly \textit{how} the system has evolved after 4 years of heat demand growth, we now compare the Explicit heat model \eqref{e:D_diaggr} and Implicit heat model \eqref{e:ldc_peak} using the central heat pump profile $\mathcal{C}$ with a COP of 2.0. The Implicit model requires the zero-heat demand model of $\tau-0$ (the 20/21 winter) to be scaled by 14.0\% so that the Peak Demand matches the Explicit model (as described in Section \ref{ss:implicit}), and so all coefficients have increased by the same amount compared to that model. In contrast, the coefficients of the Explicit model have changed in more subtle ways to match the shape of the electrical-plus-heat demand model.

\begin{figure}\centering
\subfloat[Explicit Heat Model]{\includegraphics[width=0.49\textwidth]{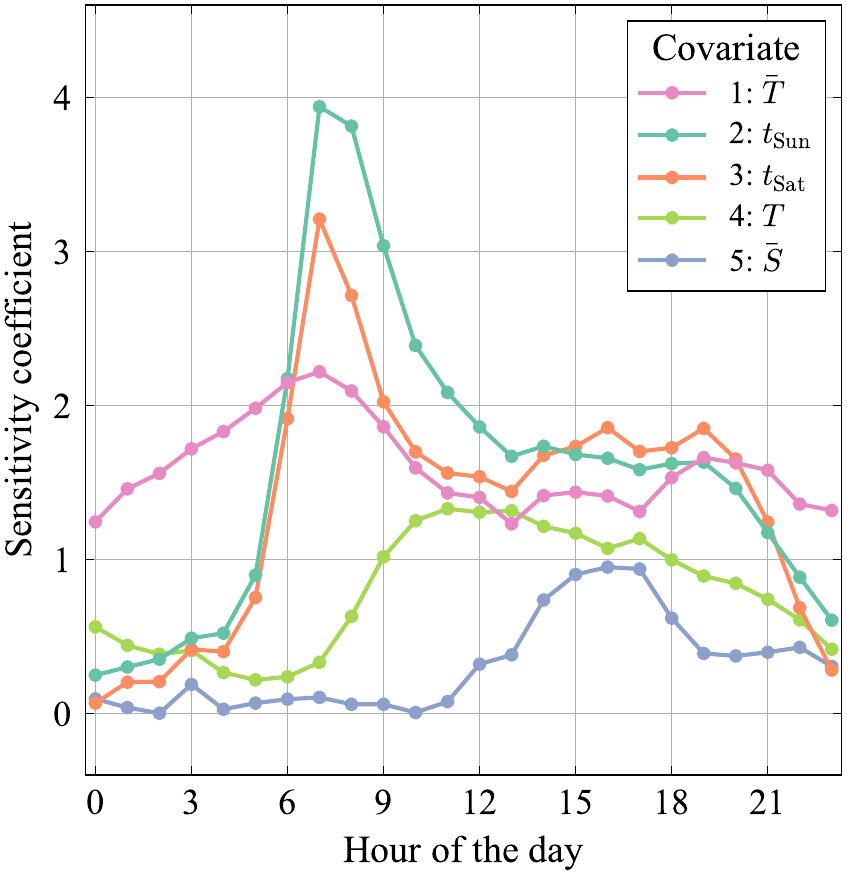}\label{f:plot_coefs_gas}}~
\subfloat[Implicit Heat Model]{\includegraphics[width=0.49\textwidth]{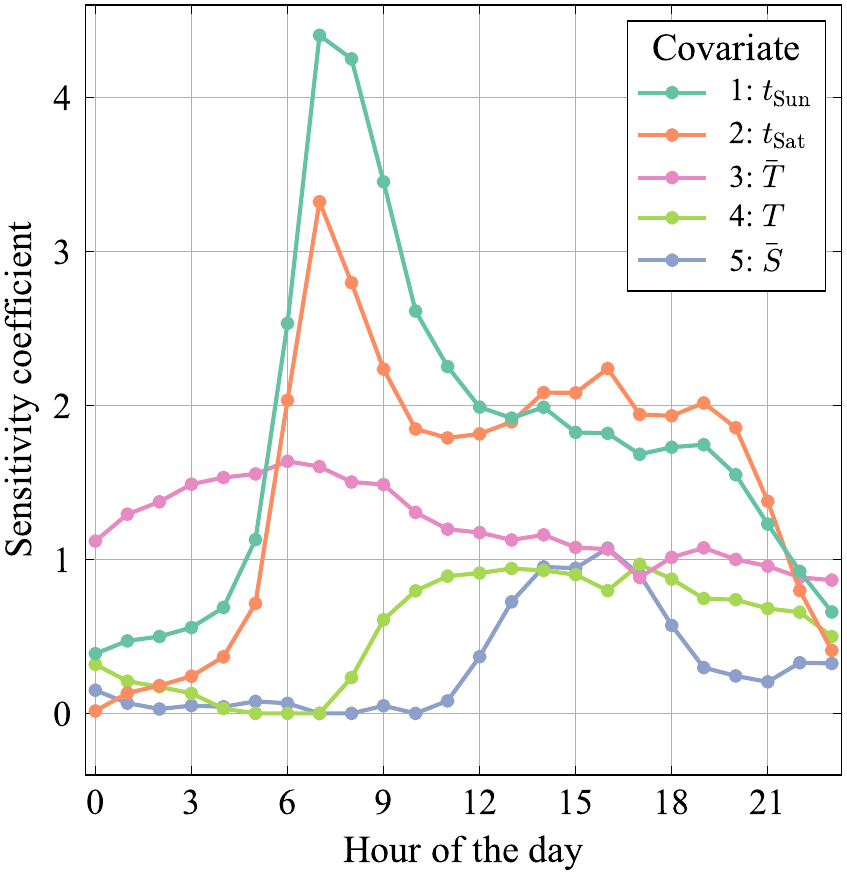}\label{f:plot_coefs_peak}}
\caption{A comparison between the Explicit and Implicit heat demand growth model using COP of 2.0 and the central HP profile $\mathcal{C}$ for $\tau -4$ delivery. The five covariates with the largest mean absolute value are ordered according to that mean value. The mean temperature sensitivity (calculated as the mean of the sum of coefficients associated with temperature variables $T,\,\bar{T}$) increases by 54\% with the Explicit model to 2.4~GW per unit of normalized temperature. In comparison, the mean temperature sensitivity only increases by 14.0\% with the Implicit heat demand model to 1.8~GW per unit of normalized temperature.}
\label{f:plt_coefs}
\end{figure}

Firstly, we consider the five sensitivity coefficient values with the largest mean absolute value across the day, as plotted in Figure \ref{f:plt_coefs}. (The covariates have been normalized to have zero mean and unit variance, and so a comparison in this way is meaningful.) It can be observed that sensitivities change significantly through the day--for example, mean 24-hour solar irradiance $\bar{S}$ is related to the demand, but with the highest sensitivities after 11am (even extending to the hours after dark).

A clear result is the change in sensitivity coefficients with respect to temperatures $T,\,\bar{T}$. The sensitivity coefficient corresponding to mean daily temperature $\bar{T}$ has increased sharply around 7am in the Explicit model, as compared to the value obtained by uniform scaling using the Implicit model. This sharp increase is caused by the morning peak seen in the central heat pump profile $\mathcal{C}$, which has a peak around this time (Figure \ref{f:plt_hp_variations}). There is also a noticeable reduction in the sensitivity of demand to weekends $t_{\mathrm{Sat}},\,t_{\mathrm{Sun}}$ as well using the Explicit model. 

The mean during the afternoon peak (1600-1900) and number of non-zero of each of the sensitivity coefficients (across all 24 hours), as calculated for the $\tau-4$ year 24/25, are also compared for the Implicit and Explicit models in Table \ref{t:regression_comparison}. The mean temperature sensitivity (across the instantaneous and 24 hour mean values $T,\,\bar{T}$) has increased by 54\% in the Explicit model, at a rate of more than three times that of the Implicit model. It can also be observed that, of the additional covariates selected from the Composite Weather Variable, the 24-hour wind chill $\bar{W}_{\mathrm{Chill}}$ is most influential, with the corresponding model coefficient more than quadrupling from 0.042 to a value of 0.205 during the peak hours.

\begin{table}
\centering

\begin{tabular}{llllll}
\toprule
Vrbl. & Explicit & Implicit & Vrbl. & Explicit & Implicit \\
\midrule
$t_{\mathrm{Sat}}$ & \cellcolor[HTML]{4073b4} 1.78 (24) & \cellcolor[HTML]{3e5ea9} 2.03 (24) & $t_{\mathrm{Tue}}$ & \cellcolor[HTML]{e5fafb} 0.0486 (7) & \cellcolor[HTML]{e1f7f9} 0.0748 (8) \\
$t_{\mathrm{Sun}}$ & \cellcolor[HTML]{4580b9} 1.62 (24) & \cellcolor[HTML]{4176b6} 1.74 (24) & $t_{\mathrm{Prd, C2}}$ & \cellcolor[HTML]{e5fafb} 0.0428 (9) & \cellcolor[HTML]{e8fcfc} 0.0177 (7) \\
$\bar{T}$ & \cellcolor[HTML]{4b8cbd} 1.48 (24) & \cellcolor[HTML]{6bb1cb} 1.01 (24) & $t_{\mathrm{Prd, S3}}$ & \cellcolor[HTML]{e8fcfc} 0.0178 (2) & \cellcolor[HTML]{e7fbfb} 0.039 (2) \\
$T$ & \cellcolor[HTML]{6ab0cb} 1.02 (24) & \cellcolor[HTML]{7bbfd0} 0.846 (21) & $t_{\mathrm{Mon}}$ & \cellcolor[HTML]{eafdfd} 0.013 (10) & \cellcolor[HTML]{e7fbfb} 0.0293 (14) \\
$\bar{S}$ & \cellcolor[HTML]{88c8d4} 0.724 (24) & \cellcolor[HTML]{89c9d5} 0.711 (21) & $\bar{T}_{\mathrm{Cold}}$ & \cellcolor[HTML]{eafdfd} 0.00895 (5) & \cellcolor[HTML]{eafdfd} 0 (1) \\
$t_{\mathrm{Fri}}$ & \cellcolor[HTML]{abdbe0} 0.456 (10) & \cellcolor[HTML]{97d1d9} 0.608 (12) & $W_{\mathrm{on}}$ & \cellcolor[HTML]{eafdfd} 0.00091 (5) & \cellcolor[HTML]{eafdfd} 0.0106 (9) \\
$t_{\mathrm{Lin}}$ & \cellcolor[HTML]{b7e1e4} 0.377 (10) & \cellcolor[HTML]{a6d8de} 0.495 (14) & $T_{\mathrm{Cold}}$ & \cellcolor[HTML]{eafdfd} 0 (9) & \cellcolor[HTML]{eafdfd} 0 (5) \\
$t_{\mathrm{Sunset}}$ & \cellcolor[HTML]{bce4e7} 0.34 (5) & \cellcolor[HTML]{a9dadf} 0.473 (15) & $W_{\mathrm{Chill}}$ & \cellcolor[HTML]{eafdfd} 0 (0) & \cellcolor[HTML]{eafdfd} 0 (0) \\
$t_{\mathrm{Prd, C1}}$ & \cellcolor[HTML]{c4e7ea} 0.283 (22) & \cellcolor[HTML]{d6f1f3} 0.154 (13) & $S$ & \cellcolor[HTML]{eafdfd} 0 (5) & \cellcolor[HTML]{eafdfd} 0 (5) \\
$\hat{E}_{\mathrm{Out}}$ & \cellcolor[HTML]{c6e8eb} 0.269 (11) & \cellcolor[HTML]{c2e6e9} 0.294 (12) & $W_{\mathrm{off}}$ & \cellcolor[HTML]{eafdfd} 0 (7) & \cellcolor[HTML]{e5fafb} 0.0404 (22) \\
$\bar{W}_{\mathrm{Chill}}$ & \cellcolor[HTML]{cfedef} 0.205 (10) & \cellcolor[HTML]{e5fafb} 0.0419 (3) & $t_{\mathrm{Prd, S1}}$ & \cellcolor[HTML]{eafdfd} 0 (0) & \cellcolor[HTML]{eafdfd} 0 (0) \\
$t_{\mathrm{Prd, S2}}$ & \cellcolor[HTML]{e3f9fa} 0.0553 (3) & \cellcolor[HTML]{d8f2f4} 0.144 (2) & $t_{\mathrm{Thu}}$ & \cellcolor[HTML]{eafdfd} 0 (9) & \cellcolor[HTML]{eafdfd} 0 (4) \\
$t_{\mathrm{Prd, C3}}$ & \cellcolor[HTML]{e5fafb} 0.0527 (1) & \cellcolor[HTML]{e1f7f9} 0.0739 (1) & $\hat{G}_{\mathrm{Out}}$ & \cellcolor[HTML]{eafdfd} 0 (0) & \cellcolor[HTML]{eafdfd} 0 (0) \\
\bottomrule
\end{tabular}
\caption{Comparison of Explicit \eqref{e:D_diaggr} and Implicit \eqref{e:ldc_peak} models for $\tau - 4$ delivery (24/25). Despite the models having the same peak demands, the coefficients differ as the Explicit model is fit using gas demand as a proxy for heat, where the Implicit model only scales existing load duration curves. Reported are the mean of the regression coefficients during the system peak (1600-1900 hours); numbers in parenthesis indicate the number of non-zero coefficients across the whole day.}\label{t:regression_comparison}

\end{table}

Interestingly, there is not enough correlation between many of the other weather-based covariates and demand to yield a non-zero response. This is potentially due to the scope of the modeling considered here as compared to the industry models from which they are derived: we have only considered the coldest months, where approaches such as the Composite Weather Variable are designed to model year-round weather sensitivities. For example, covariates such as the cold-spell uptick $T_{\mathrm{Cold}}$ could account for the fact that demand-temperature sensitivities are known to be different in winter and summer in the GB system \cite{thornton2016role}.

\subsection{Bias in ACTS Calculations from Implicit Heat Demand Modelling}\label{ss:bias}

Having compared the sensitivities of the Implicit and Explicit models, we now consider how the ACTS capacity changes for each of these models through to $\tau-4$ delivery. Table \ref{t:cap2sec} reports the Additional Capacity to Secure for each of the five years to $\tau-4$ for the central heat pump profile $\mathcal{C}$ and COP of 2.0. Whilst the underlying electrical demand $E$ does not change, the generation fleet $X$ does, particularly from $\tau-3$ to $\tau-4$, as there is significant levels of decommissioning of legacy plant expected during this period (Figure \ref{f:pltGenBoxplots}). It can be seen that as the level of electrical heat demand $H$ increases, ACTS calculated by Explicit and Implicit system models drifts apart, with increasing over-procurement Bias \eqref{e:bias}. By delivery at the $\tau-4$ time period, the Bias is such that 0.79~GW additional ACTS is required for the Implicit model as compared to the Explicit model. By comparison, the Demand Curves used in the GB capacity market $\tau-1$ and $\tau-4$ auctions have had a width of 2 GW or less for the past two years \cite{beis2020letter}. This Bias could lead to systematic over-investment--if at $\tau-1$ there is a consistent over-procurement of 0.21~GW over 10 years at the net Cost of New Entry of £49m per GW-yr \cite{beis2020letter}, this equates to an over-spend of £103m.

\begin{table}
\centering

\begin{tabular}{lll}
\toprule
Delivery year & Explicit, GW & Implicit, GW \\
\midrule
$\tau-$0 (20/21) & $-4.36$ & $-4.36$ \\
$\tau-$1 (21/22) & $-3.75$ & $-3.54$ \\
$\tau-$2 (22/23) & $-2.02$ & $-1.6$ \\
$\tau-$3 (23/24) & $2.02$ & $2.62$ \\
$\tau-$4 (24/25) & $6.26$ & $7.05$ \\
\bottomrule
\end{tabular}
\caption{Comparison of Additional Capacity to Secure, comparing the Explicit model \eqref{e:margin}, \eqref{e:D_diaggr} against the Implicit model \eqref{e:margin}, \eqref{e:ldc_peak}, using the central heat pump profile $\mathcal{C}$ and a COP of 2.0.}\label{t:cap2sec}

\end{table}

A similar result is found for the other two heat pump profiles, as given in Table \ref{t:tbl_final_comparison}. The ACTS is biased by 0.71~GW for the flat profile $\mathcal{F}$, a similar amount to the central profile $\mathcal{C}$. The peak profile $\mathcal{P}$ shows an even larger change of more than 2.3~GW, but that is perhaps unsurprising giving the extreme evening peak in the evening of that profile (Figure \ref{f:plt_hp_variations}). 

\begin{table}
\centering

\begin{tabular}{llllllllll}
\toprule
& \multicolumn{3}{c}{$\mathcal{C}$ (Love et al.)} & \multicolumn{3}{c}{$\mathcal{F}$ (Eyre et al.)} & \multicolumn{3}{c}{$\mathcal{P}$ (Sansom)} \\
    \cmidrule(l{0.6em}r{0.9em}){2-4} \cmidrule(l{0.6em}r{0.9em}){5-7} \cmidrule(l{0.6em}r{0.9em}){8-10}
    & Ex. & Im. & Bias & Ex. & Im. & Bias & Ex. & Im. & Bias\\
    \midrule
ACTS, GW$:$ & 6.26 & 7.05 & 0.79 & 5.43 & 6.14 & 0.71 & 12.82 & 15.15 & 2.33 \\
\bottomrule
\end{tabular}
\caption{Comparison of the Explicit (Ex.) and Implicit (Im.) ACTS for $\tau - 4$ delivery (24/25) and corresponding Bias \eqref{e:bias} for each of the three heat pump profiles, assuming a system COP of~2.0.}\label{t:tbl_final_comparison}

\end{table}

\begin{figure}\centering
\includegraphics[width=0.86\textwidth]{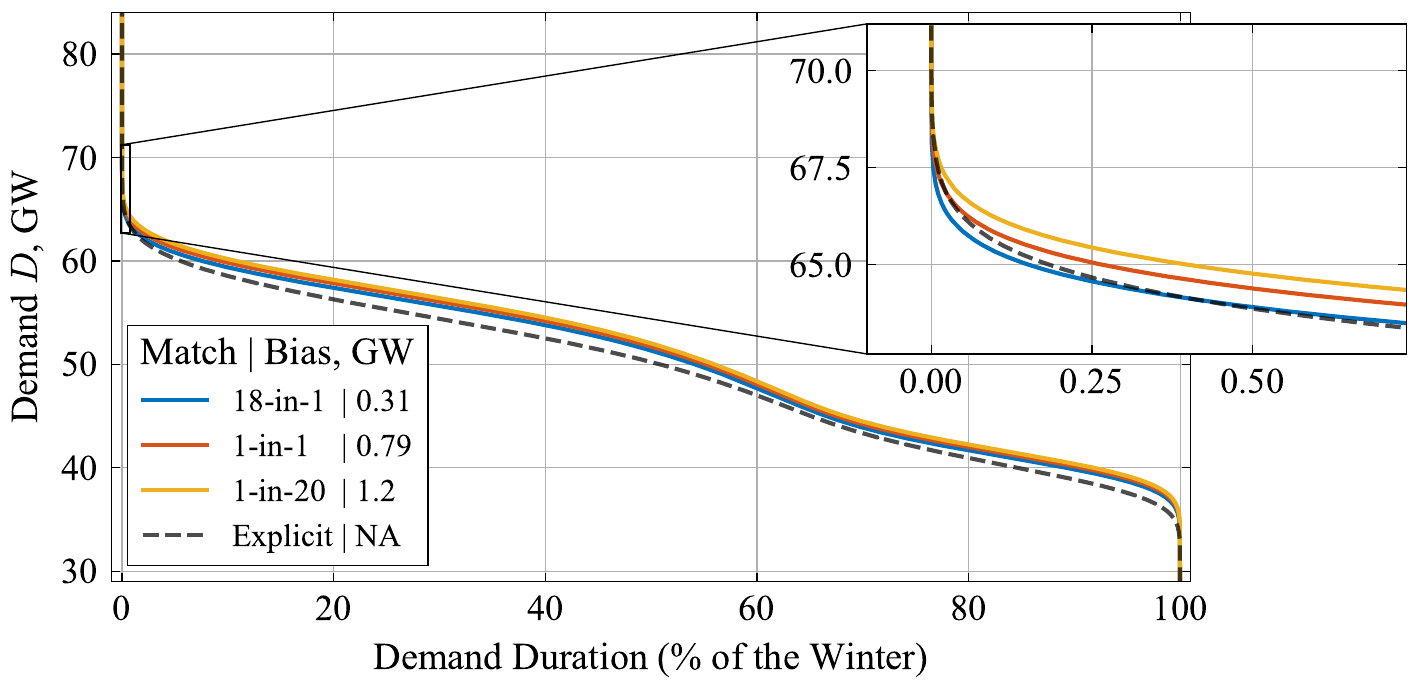}
\caption{Comparison between the load duration curves of the $\tau-4$ Explicit model \eqref{e:D_diaggr} using central HP profile $\mathcal{C}$ and COP of 2.0, in comparison to the equivalent Implicit models \eqref{e:ldc_peak}, matching the eighteen-per-year, once-per-year and one-in-twenty median Peak Demands. The median demand is several GW higher for the Implicit models--it is only at high demand levels that the distribution of the Implicit models become close to that of the Explicit models (inset).}\label{f:plt_bias_ldcs}
\end{figure}

To consider why this bias is introduced, we plot the load duration curve (LDC) of the demand $D$ of the Explicit model for $\tau-4$ in Figure \ref{f:plt_bias_ldcs}, with three different Implicit models fit. This first is a model to fit the once-per-year peak demand (as in \eqref{e:def_peak}); the other two models fit the one-in-twenty-year median peak demand, and the eighteen-per-year median peak demand (i.e., the weekly median peak demand). As the Explicit and Implicit models have difference distributions, this results in different demand profiles $D$, and subsequently different levels of Bias. It can be seen that the value of demand $D$ at 50\% duration is greater in the Implicit than Explicit models. This implies that the rate of increase of the Peak Demand of $D$ is greater than that of the median demand of $D$--the Implicit scaling coefficient $k_{\mathrm{Peak}}$ must over-compensate for the increased Peak Demand, so overshoots the quantiles lower down the LDC.

\subsection{Sensitivity Analysis: Varying Heat Profiles and System Coefficient of Performance}\label{ss:variability}

As mentioned previously, sensitivity analysis is an integral part of resource adequacy planning and capacity markets, as there are many uncertainties for which it is difficult to assign probabilities accurately. In this final section we consider how the variability of ACTS capacity changes, considering first increased meteorological uncertainty (from increased demand-weather sensitivity) and secondly in uncertainty to heat pump profiles $h$ and system-wide coefficient of performance $k_{\mathrm{COP}}$. The goal is to compare the additional variability compared to existing variations considered by industry, and put into perspective the bias observed in the previous section.

\subsubsection{Meteorological Sensitivity at $\tau - 4$ Delivery}\label{sss:weather}

To consider clearly how the meteorological sensitivity has changed, the first result we consider is how the $\tau - 4$ delivery compares for models with and without heat pumps. Results are given in Table \ref{t:acts_warm_cold} for the central profile $\mathcal{C}$ and COP of 2.0. It is found that the ACTS increases from $-0.72$ (i.e., a small level of oversupply) to under-supply of 6.26~GW, an increase in ACTS of 7.0~GW. This implies an increase in demand of 1.75 kW per heat pump at peak. This is remarkably close to the 1.7 kW per heat pump estimated by \cite{love2017addition} and well within the range of 1.2 to 2.6 kW reported in \cite{cooper2016detailed}.

\begin{table}
\centering

\begin{tabular}{llllll}
\toprule
\multirow{2}{*}{Model} & \multirow{2}{*}{\shortstack{$1990-2019$\\climate ACTS}} & \multicolumn{2}{c}{Warm weather (13/14)} & \multicolumn{2}{c}{Cold weather (10/11)}\\
     \cmidrule(l{0.6em}r{0.9em}){3-4} \cmidrule(l{0.6em}r{0.9em}){5-6}
     & & ACTS & Change & ACTS & Change \\\midrule
No HPs & -0.72 & -2.61 & -1.89 & 1.01 & 1.73 \\
With HPs & 6.26 & 3.23 & -3.03 & 9.01 & 2.75 \\
\bottomrule
\end{tabular}
\caption{Changes in Additional Capacity to Secure (in GW) for $\tau - 4$ delivery (24/25), comparing capacity requirements for the long-term climate against warm (13/14) and cold (10/11) years. The model with heat pumps (HPs) is the Explicit model.}\label{t:acts_warm_cold}

\end{table}

Of particular interest for this work, however, is how changing the winter conditions changes the ACTS compared to the use of a long-term average. The 10/11 and 13/14 winters are chosen first for investigation as these have been identified in prior works as being of particular importance for the determination of peaking capacity in the GB system \cite{hilbers2019importance}. These winters have mean hourly temperatures $T$ of 7.98\degC, 10.13\degC~(the long-term mean and standard deviation of $T$ are 9.46\degC, 3.4\degC, respectively); the mean offshore capacity factors $W_{\mathrm{Off}}$ are 0.39 and 0.55 (with long-term mean and standard deviation of 0.51, 0.28 respectively). The results shows an almost-symmetric change in the required capacity factor close to 1.8~GW for those particular years with no heat pumps (Table \ref{t:acts_warm_cold}). As the system meteorological sensitivity increases with heat pump numbers, however, the width of spread for those anomalous years has increased by almost 50\%, subsequently covering the range from $-3.03$ to 2.75~GW. 

\begin{figure}\centering
\includegraphics[width=0.99\textwidth]{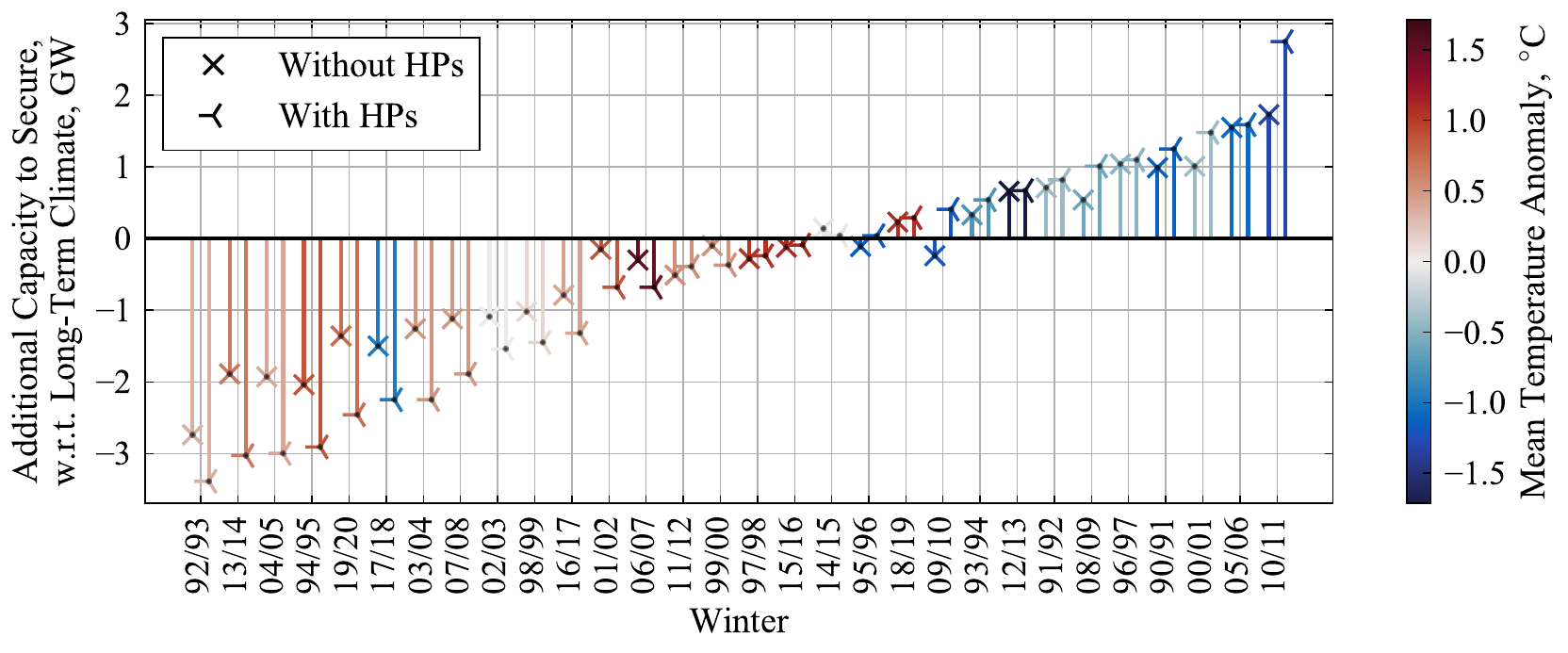}
\caption{Changes in ACTS, comparing values if all years have the same weather against the 30-year climate, calculated for the $\tau-4$ delivery (24/25). The addition of heat pumps leads to an increased range of ACTS with respect to the long-term winter climate, due to increased weather sensitivity; although, it is clear that the mean temperature anomaly only partially explains the change in ACTS of individual winters.}\label{f:plot_weather_sns}
\end{figure}

Furthermore, Figure \ref{f:plot_weather_sns} plots the changes in ACTS for all winters, again comparing the model with and without heat pumps. In this figure, the colour of the points indicates the value of the mean temperature anomaly for that winter. From this figure, it is first observed that 10/11 and 13/14 winters are close to the extremes for the ACTS spread at $\tau-4$ delivery, as expected. The use of several decades of climate data allows for a much deeper understanding of the variability of these estimates, and allows for the severity of individual winters to be compared.

It is interesting to note from this figure, however, that it is not only the mean temperature that determines the change in ACTS--the Pearson correlation coefficient between the mean temperature anomaly and the change in ACTS is $r=-0.55$. For example, the 17/18 winter is cooler than the long-term average but requires less generating capacity than the long-term climate, as the very cold temperatures from that winter do not occur at points of high net demand. This is also possible to observe in winters such as 06/07 and 12/13--whilst these are the extreme years in terms of mean temperature anomaly, they do not lead the greatest changes in ACTS. As both wind generation and heat pump demand increases, the sensitivity of the system to wind (through renewables) and temperature (through demand) will continue to evolve, and so the meteorological conditions of greatest importance to explain power system variability will also continue to change.

\subsubsection{Impacts of Changing System Performance Factors}

\begin{figure}\centering
\includegraphics[width=0.88\textwidth]{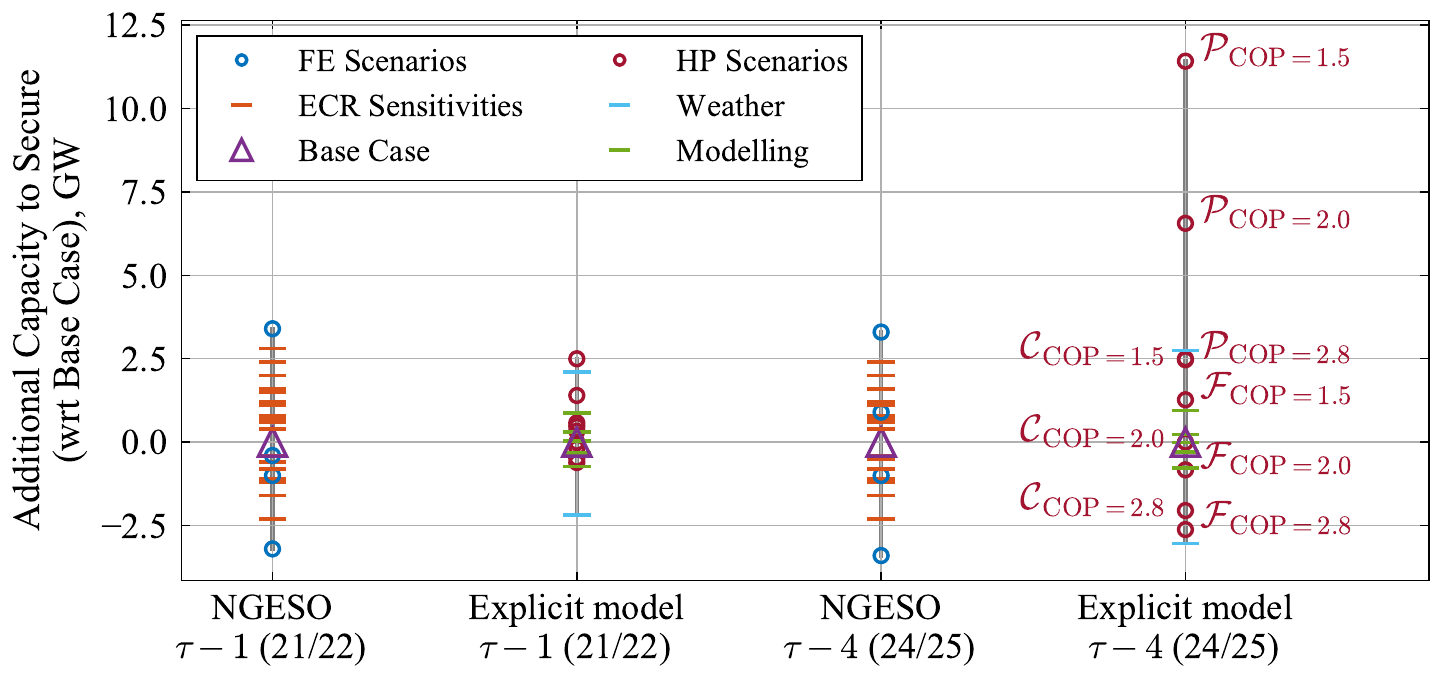}
\caption{Sensitivity analysis for $\tau-1$ and $\tau-4$ cases, comparing the Explicit model's range of ACTS calculated over a range of scenarios and sensitivities with the equivalent from NGESO's most recent electricity capacity market reports. Uncertainties in Weather and Modelling for Explicit modelling are considered for the central HP profile $\mathcal{C}$ with COP of 2.0.}\label{f:plt_hp_snstvty}
\end{figure}

The ACTS for each combination of heat pump profile $h$ and coefficient of performance $k_{\mathrm{COP}}$ is now considered. Figure \ref{f:plt_hp_snstvty} plots these sensitivities against Weather and Modelling sensitivities, as well as against the range of scenarios and sensitivities considered within the GB capacity market \cite{ngeso2020ecr}. Weather sensitivities are based on the cold/warm winters of 10/11 and 13/14 (as in Section \ref{sss:weather}), whilst the Modelling uncertainties represent the ACTS based on the five cross-validation models (i.e., five models fit with the optimal $\alpha^{*}$ in \eqref{e:theta_lasso}, but with the data of one winter not included). NGESO's FE scenarios consider a range of supply and demand pathways up to 2050, whilst ECR Sensitivities consists of (amongst others) over- and under-delivery of the capacity market, unusually cold or warm weather, and over/under estimation of nominal peak demands.

The Explicit modelling uncertainty around $\tau-1$ delivery has a RoCS of 4.7~GW, increasing to 14.5~GW for $\tau-4$. In comparison, the current industry (NGESO) sensitivities and scenarios for $\tau-1$ and $\tau-4$ remain roughly equivalent in their Range of Capacity to Secure, with total RoCS of 6.6~GW and 6.7~GW respectively. In other words, the uncertainty due to the addition of one million heat pumps per year, given the three heat pump profiles $\mathcal{C},\,\mathcal{F},\,\mathcal{P},\,$ and COPs from 1.5-2.8 is more than double that of all sensitivities considered in the capacity market report for 2020, even without considering variations in actual installed number of installed heat pumps $n_{H}$. This huge range in possible outcomes for high penetrations of heat pumps has been noted before \cite{cooper2016detailed}, although to our knowledge has not been identified as an issue in the context of capacity markets.

The unrestricted profile $\mathcal{P}$ of Sansom is considered pessimistic \cite{watson2019decarbonising}, with a peak demand much greater than the other profiles. In the worst case (with system COP of 1.5), four million heat pumps would each add 4.6 kW to the peak. In terms of peak \textit{heat} demand per dwelling, this is not unreasonable--measured data analysed in \cite{watson2019decarbonising} shows a heat demand peak of 8 kW per dwelling at the coldest temperatures, and so the value calculated in this work would correspond to an equivalent COP of 1.7. However, heat pumps tend to have very different load profiles to gas boilers, as equivalent devices of the same volume and footprint are much lower power, generally resulting in much flatter profiles.

Nevertheless, even with the unrestricted profile $\mathcal{P}$ excluded, the uncertainty in peak demand increase using only profiles $\mathcal{C},\,\mathcal{F}$ shows a Range of Capacity to Secure of 5.05~GW, 75\% of the $\tau-4$ uncertainty of NGESO (again, even without considering variation in the number of actual heat pumps installed or heat pump DSR). These numbers are comparable with white-box modelling estimates such as \cite{cooper2016detailed}, which (by consideration only of the reported range of possible peak demands of heat pumps of between 1.2 kW and 2.6 kW) would show a range of 5.1~GW with four million heat pumps. If this uncertainty is added to the existing variability in industry forecasts (as the FE scenario only consider very low levels of annual heat pump installations), the RoCS would increase to more than 10 GW.

\subsubsection{Discussion: Bias, Variability and Impacts on Social Welfare}

To close this section, we briefly compare bias and variability from sensitivities and scenarios in the context of the capacity market. Where possible, bias should clearly be eliminated: if we consistently over- or under-estimate the true capacity to secure, then in the long-run consumers will pay for unnecessary capacity (the Cost of New Entry in the GB capacity market is £49m per GW-yr \cite{ngeso2020ecr}). Changes in demand-weather sensitivity lead to uncertainty that is largely irreducible, and so it is sensible to ensure this sort of risk can be communicated to the market. Whilst uncertainties around the number of heat pumps installed will likely remain, uncertainties pertaining to future heat demand curves could be diminished significantly.

Given this conclusion, system operators need to continuously review the \textit{most extreme} scenarios, as well as understanding estimates of possible bias in their analysis. For example, if the peaking profile $\mathcal{P}$ could confidently be rejected, then this would deliver immediate benefits, as the target capacity to secure can be reduced to more modest levels. Similarly, if a system operator considers the Explicit heat model as the most accurate modelling approach, then using this model would eliminate Bias (as well as any potential of perceived bias), thereby providing both accurate results and modelling transparency.

\section{Conclusions}\label{s:conclusions}

This paper considered the impacts of electrification of heat on system adequacy, with a particular focus on the changes to the Additional Capacity to Secure driven by both scenario and meteorological uncertainties. The addition of one million heat pumps annually to the GB system is considered as a case study, which has a very large impact on the capacity required to reach system security standards.

In particular, it is shown that demand-weather sensitivities could increase by 50\% for $\tau-4$ delivery, with similar increase in the Range of Capacity to Secure required when net demand is conditioned on individual weather years. The proposed Explicit heat demand model has advantages in terms of transparency, with the heterogeneous differences between space heating and electrical demand growth clearly demarcated. By utilising Lasso-Regularized linear regression methods, a wide range of weather-based covariates can be studied without the problems of over-fitting that occur using the standard Least-Squares approach.

It is shown that failure to disaggregate demand separately into heat and electrical demands leads to a bias of 0.79~GW in capacity to secure when compared to results determined using the legacy approach of linearly scaling load duration curves. Over the course of 10 years, it is demonstrated this could result in over-procurement of more than £100m. Uncertainty due to heat pump demand could be greater than all currently considered scenarios that NGESO considers, although further research to narrow this uncertainty could prove fruitful. Reducing this uncertainty, in-turn, would lead to benefits in terms of reducing wasteful over-procurement of peaking capacity, although this needs to be done conservatively to ameliorate dangers of costly lost load.

There are many ways in which the electrification of heating demand can impact energy system security. Amongst others, interactions with large-scale energy storage introduces time coupling which adds complexities (whether from standalone electrical, thermal, or electric vehicle storage). Increased interconnection between regions which each have high levels of space heating could increase further challenges around coincidence of peaks between markets which had previously been considered largely independent. It is concluded that modelling rapid electrification of heat in such interconnected systems will require multi-region, time-sequential modelling for the accurate determination of capacity requirements.

\subsection*{Acknowledgments}
The authors are grateful for the help of Archie Corliss and Katherine Daman of National Grid ESO, for their time discussing, and provision of data on, customer demand management. The work was funded by the Engineering and Physical Sciences Research Council through grant no. EP/S00078X/1 (Supergen Energy Networks hub 2018). H. Bloomfield and D. Greenwood are funded by the Supergen Energy Network's CLEARHEADS flex project. S. Sheehy is funded by an EPSRC studentship.

\bibliography{refs_aeDec2020}

\begin{thebibliography}{10}
\expandafter\ifx\csname url\endcsname\relax
  \def\url#1{\texttt{#1}}\fi
\expandafter\ifx\csname urlprefix\endcsname\relax\def\urlprefix{URL }\fi
\expandafter\ifx\csname href\endcsname\relax
  \def\href#1#2{#2} \def\path#1{#1}\fi

\bibitem{eggimann2019high}
S.~Eggimann, J.~W. Hall, N.~Eyre, A high-resolution spatio-temporal energy
  demand simulation to explore the potential of heating demand side management
  with large-scale heat pump diffusion, Applied Energy 236 (2019) 997--1010.

\bibitem{eggimann2020weather}
S.~Eggimann, W.~Usher, N.~Eyre, J.~W. Hall, How weather affects energy demand
  variability in the transition towards sustainable heating, Energy 195 (2020)
  116947.

\bibitem{staffell2018increasing}
I.~Staffell, S.~Pfenninger, The increasing impact of weather on electricity
  supply and demand, Energy 145 (2018) 65--78.

\bibitem{wilson2013historical}
I.~G. Wilson, A.~J. Rennie, Y.~Ding, P.~C. Eames, P.~J. Hall, N.~J. Kelly,
  Historical daily gas and electrical energy flows through {Great Britain}'s
  transmission networks and the decarbonisation of domestic heat, Energy Policy
  61 (2013) 301--305.

\bibitem{bloomfield2016quantifying}
H.~C. Bloomfield, D.~J. Brayshaw, L.~C. Shaffrey, P.~J. Coker, H.~Thornton,
  Quantifying the increasing sensitivity of power systems to climate
  variability, Environmental Research Letters 11~(12) (2016) 124025.

\bibitem{ngeso2017ecr}
{National Grid ESO}, {National Grid EMR}: Electricity capacity report 2017
  (2017).

\bibitem{soder2020review}
L.~S{\"o}der, E.~T{\'o}masson, A.~Estanqueiro, D.~Flynn, B.-M. Hodge,
  J.~Kiviluoma, M.~Korp{\aa}s, E.~Neau, A.~Couto, D.~Pudjianto, et~al., Review
  of wind generation within adequacy calculations and capacity markets for
  different power systems, Renewable and Sustainable Energy Reviews 119 (2020)
  109540.

\bibitem{cole2020considerations}
W.~Cole, D.~Greer, J.~Ho, R.~Margolis, Considerations for maintaining resource
  adequacy of electricity systems with high penetrations of {PV} and storage,
  Applied Energy 279 (2020) 115795.

\bibitem{dent2016capacity}
C.~Dent, R.~Sioshansi, J.~Reinhart, A.~Wilson, S.~Zachary, M.~Lynch,
  C.~Bothwell, C.~Steele, Capacity value of solar power: Report of the {IEEE
  PES} task force on capacity value of solar power, PMAPS, Beijing, China
  (2016).

\bibitem{edwards2017assessing}
G.~Edwards, S.~Sheehy, C.~J. Dent, M.~C. Troffaes, Assessing the contribution
  of nightly rechargeable grid-scale storage to generation capacity adequacy,
  Sustainable Energy, Grids and Networks 12 (2017) 69--81.

\bibitem{nolting2020can}
L.~Nolting, A.~Praktiknjo, Can we phase-out all of them? probabilistic
  assessments of security of electricity supply for the german case, Applied
  Energy 263 (2020) 114704.

\bibitem{khan2018demand}
A.~S.~M. Khan, R.~A. Verzijlbergh, O.~C. Sakinci, L.~J. De~Vries, How do demand
  response and electrical energy storage affect (the need for) a capacity
  market?, Applied Energy 214 (2018) 39--62.

\bibitem{lynch2019impacts}
M.~{\'A}. Lynch, S.~Nolan, M.~T. Devine, M.~O’Malley, The impacts of demand
  response participation in capacity markets, Applied Energy 250 (2019)
  444--451.

\bibitem{zachary2019integration}
S.~Zachary, A.~Wilson, C.~Dent, The integration of variable generation and
  storage into electricity capacity markets, whitepaper. arXiv preprint
  arXiv:1907.05973 (2019).

\bibitem{pte2020report}
D.~Bunn, G.~Doyle, N.~Jenkins, F.~Kelly, L.~Waters, {Panel of Technical
  Experts}: Report on the {National Grid ESO} electricity capacity report 2020,
  \url{https://www.gov.uk/government/publications/national-grid-eso-electricity-capacity-report-2020-findings-of\\
  -the-panel-of-technical-experts}, accessed 8/2/21 (July 2020).

\bibitem{bossmann2015shape}
T.~Bo{\ss}mann, I.~Staffell, The shape of future electricity demand: Exploring
  load curves in 2050s {Germany} and {Britain}, Energy 90 (2015) 1317--1333.

\bibitem{gross2019path}
R.~Gross, R.~Hanna, Path dependency in provision of domestic heating, Nature
  Energy 4~(5) (2019) 358--364.

\bibitem{hall2016review}
L.~M. Hall, A.~R. Buckley, A review of energy systems models in the {UK}:
  Prevalent usage and categorisation, Applied Energy 169 (2016) 607--628.

\bibitem{scamman2020heat}
D.~Scamman, B.~Solano-Rodr{\'\i}guez, S.~Pye, L.~F. Chiu, A.~Z. Smith,
  T.~Gallo~Cassarino, M.~Barrett, R.~Lowe, Heat decarbonisation modelling
  approaches in the {UK}: An energy system architecture perspective, Energies
  13~(8) (2020) 1869.

\bibitem{clegg2019integrated}
S.~Clegg, P.~Mancarella, Integrated electricity-heat-gas modelling and
  assessment, with applications to the {Great Britain} system. {Part I}:
  High-resolution spatial and temporal heat demand modelling, Energy 184 (2019)
  180--190.

\bibitem{clegg2019integrated2}
S.~Clegg, P.~Mancarella, Integrated electricity-heat-gas modelling and
  assessment, with applications to the {Great Britain} system. {Part II}:
  Transmission network analysis and low carbon technology and resilience case
  studies, Energy 184 (2019) 191--203.

\bibitem{cooper2016detailed}
S.~J. Cooper, G.~P. Hammond, M.~C. McManus, D.~Pudjianto, Detailed simulation
  of electrical demands due to nationwide adoption of heat pumps, taking
  account of renewable generation and mitigation, IET Renewable Power
  Generation 10~(3) (2016) 380--387.

\bibitem{eyre2015uncertainties}
N.~Eyre, P.~Baruah, Uncertainties in future energy demand in {UK} residential
  heating, Energy Policy 87 (2015) 641--653.

\bibitem{hyndman2009density}
R.~J. Hyndman, S.~Fan, Density forecasting for long-term peak electricity
  demand, IEEE Transactions on Power Systems 25~(2) (2009) 1142--1153.

\bibitem{li2019use}
Y.~Li, B.~Jones, The use of extreme value theory for forecasting long-term
  substation maximum electricity demand, IEEE Transactions on Power Systems
  35~(1) (2019) 128--139.

\bibitem{khuntia2016forecasting}
S.~R. Khuntia, J.~L. Rueda, M.~A. van~der Meijden, Forecasting the load of
  electrical power systems in mid-and long-term horizons: a review, IET
  Generation, Transmission \& Distribution 10~(16) (2016) 3971--3977.

\bibitem{deakin2020calculations}
M.~Deakin, S.~Sheehy, D.~Greenwood, S.~Walker, P.~Taylor, Calculations of
  system adequacy considering heat transition pathways, in: 2020 IEEE
  International Conference on Probabilistic Methods Applied to Power Systems
  (PMAPS), IEEE, 2020, pp. 1--6.

\bibitem{ofgem2013electricity}
P.~Ochoa, Electricity capacity assessment report 2013, Tech. rep., {OFGEM}
  (2013).

\bibitem{love2017addition}
J.~Love, A.~Z. Smith, S.~Watson, E.~Oikonomou, A.~Summerfield, C.~Gleeson,
  P.~Biddulph, L.~F. Chiu, J.~Wingfield, C.~Martin, et~al., The addition of
  heat pump electricity load profiles to {GB} electricity demand: Evidence from
  a heat pump field trial, Applied Energy 204 (2017) 332--342.

\bibitem{wilson2018use}
A.~L. Wilson, S.~Zachary, C.~J. Dent, Use of meteorological data for improved
  estimation of risk in capacity adequacy studies, in: 2018 IEEE International
  Conference on Probabilistic Methods Applied to Power Systems (PMAPS), IEEE,
  2018, pp. 1--6.

\bibitem{ng2021acs}
{National Grid ESO}, Average cold spell methodology,
  \url{https://www.emrdeliverybody.com/Capacity%20Markets%20Document%20Library/NGESO%20ACS%20Methodology.pdf}
  (Accessed Jan. 2021).

\bibitem{cramton2005capacity}
P.~Cramton, S.~Stoft, A capacity market that makes sense, The Electricity
  Journal 18~(7) (2005) 43--54.

\bibitem{cramton2017electricity}
P.~Cramton, Electricity market design, Oxford Review of Economic Policy 33~(4)
  (2017) 589--612.

\bibitem{newbery2016missing}
D.~Newbery, Missing money and missing markets: Reliability, capacity auctions
  and interconnectors, Energy Policy 94 (2016) 401--410.

\bibitem{hawker2016capacity}
G.~Hawker, K.~Bell, S.~Gill, Capacity markets and the {EU} target model--a
  {Great Britain} case study, CIGRE (2016).

\bibitem{clegg2016assessment}
S.~Clegg, P.~Mancarella, Assessment of the impact of heating on integrated gas
  and electrical network flexibility, in: 2016 Power Systems Computation
  Conference (PSCC), IEEE, 2016, pp. 1--7.

\bibitem{eurostat2018questionnaire}
Eurostat, Questionnaire for statistics on final energy consumption in
  households, \url{https://ec.europa.eu/eurostat/web/energy/data}, accessed
  5/12/20 (2018).

\bibitem{ng2016gas}
{National Grid Gas}, Gas demand forecasting methodology,
  \url{https://www.nationalgrid.com/sites/default/files/documents/8589937808-Gas%20Demand%20Forecasting%20Methodology.pdf},
  accessed 5/12/20 (November 2020).

\bibitem{ruhnau2019time}
O.~Ruhnau, L.~Hirth, A.~Praktiknjo, Time series of heat demand and heat pump
  efficiency for energy system modeling, Scientific Data 6~(1) (2019) 1--10.

\bibitem{ngeso2019ecr}
{National Grid ESO}, Electricity capacity report 2019,
  \url{https://www.emrdeliverybody.com/CM/Guidance.aspx}, accessed 8/2/21
  (2019).

\bibitem{de2015seasonal}
M.~De~Felice, A.~Alessandri, F.~Catalano, Seasonal climate forecasts for
  medium-term electricity demand forecasting, Applied Energy 137 (2015)
  435--444.

\bibitem{hong2016probabilistic}
T.~Hong, P.~Pinson, S.~Fan, H.~Zareipour, A.~Troccoli, R.~J. Hyndman,
  Probabilistic energy forecasting: Global energy forecasting competition 2014
  and beyond, International Journal of Forecasting 32~(3) (2016) 896 -- 913.

\bibitem{hilbers2019importance}
A.~P. Hilbers, D.~J. Brayshaw, A.~Gandy, Importance subsampling: improving
  power system planning under climate-based uncertainty, Applied Energy 251
  (2019) 113114.

\bibitem{bloomfield2020characterizing}
H.~C. Bloomfield, D.~J. Brayshaw, A.~J. Charlton-Perez, Characterizing the
  winter meteorological drivers of the {European} electricity system using
  targeted circulation types, Meteorological Applications 27~(1) (2020).

\bibitem{xoserve2014autumn}
Xoserve, Autumn 2014 composite weather variable ({CWV}) review - proposed
  approach, \url{https://tinyurl.com/uebm5sx}, accessed 8/2/21 (2014).

\bibitem{desc2019review}
{Xoserve Demand Estimation Sub Committee}, Review of {CWV} optimisation,
  \url{https://gasgov-mst-files.s3.eu-west-1.amazonaws.com/s3fs-public/ggf/2019-09/2.1_Review%20of%20CWV%20Optimisation_071019.pdf},
  accessed 5/12/20 (October 2019).

\bibitem{staffell2016using}
I.~Staffell, S.~Pfenninger, Using bias-corrected reanalysis to simulate current
  and future wind power output, Energy 114 (2016) 1224--1239.

\bibitem{bloomfield2018changing}
H.~Bloomfield, D.~J. Brayshaw, L.~Shaffrey, P.~J. Coker, H.~E. Thornton, The
  changing sensitivity of power systems to meteorological drivers: a case study
  of {Great Britain}, Environmental Research Letters 13~(5) (2018) 054028.

\bibitem{cannon2015using}
D.~J. Cannon, D.~J. Brayshaw, J.~Methven, P.~J. Coker, D.~Lenaghan, Using
  reanalysis data to quantify extreme wind power generation statistics: A 33
  year case study in great britain, Renewable Energy 75 (2015) 767--778.

\bibitem{bloomfield2020merra}
H.~C. Bloomfield, D.~J. Brayshaw, A.~Charlton-Perez, {MERRA2} derived time
  series of {European} country-aggregate electricity demand, wind power
  generation and solar power generation. {(University of Reading dataset.)},
  DOI: 10.17864/1947.239 (2020).

\bibitem{gelaro2017modern}
R.~Gelaro, W.~McCarty, M.~J. Su{\'a}rez, R.~Todling, A.~Molod, L.~Takacs, C.~A.
  Randles, A.~Darmenov, M.~G. Bosilovich, R.~Reichle, et~al., The modern-era
  retrospective analysis for research and applications, version 2 (merra-2),
  Journal of Climate 30~(14) (2017) 5419--5454.

\bibitem{windpower2020}
{The Wind Power: Wind Energy Market Intelligence}, \url{thewindpower.net},
  accessed 8/2/21 (2020).

\bibitem{ngeso2019derating}
{National Grid ESO}, De-rating factor methodology for renewables participation
  in the capacity market,
  \url{https://www.emrdeliverybody.com/Capacity%20Markets%20Document%20Library/EMR%20DB%20Consultation%20response%20-%20De-rating%20Factor%20Methodology%20for%20Renewables%20Participation%20in%20the%20CM.pdf},
  accessed 8/2/21 (2019).

\bibitem{Rohatgi2020}
A.~Rohatgi, Webplotdigitizer: Version 4.4,
  \url{https://automeris.io/WebPlotDigitizer} (2020).

\bibitem{entsoe2020transparency}
{ENTSOe}, The entsoe transparency platform,
  \url{https://transparency.entsoe.eu/}, accessed 8/2/21 (2020).

\bibitem{desc2019seasonal}
{Xoserve Demand Estimation Sub Committee}, Seasonal normal review 2020: Review
  of seasonal normal basis ({SNCWV}) – {Part 1},
  \url{https://gasgov-mst-files.s3.eu-west-1.amazonaws.com/s3fs-public/ggf/2019-11/2.1_Review%20of%20Seasonal%20Normal%20Basis_SNCWV%201of2.pdf},
  accessed 28/12/20 (December 2019).

\bibitem{hastie2009elements}
T.~Hastie, R.~Tibshirani, J.~Friedman, The elements of statistical learning:
  data mining, inference, and prediction, Springer Science \& Business Media,
  2009.

\bibitem{pedregosa2011scikitLearn}
F.~Pedregosa, G.~Varoquaux, A.~Gramfort, V.~Michel, B.~Thirion, O.~Grisel,
  M.~Blondel, P.~Prettenhofer, R.~Weiss, V.~Dubourg, J.~Vanderplas, A.~Passos,
  D.~Cournapeau, M.~Brucher, M.~Perrot, E.~Duchesnay, Scikit-learn: Machine
  learning in {P}ython, Journal of Machine Learning Research 12 (2011)
  2825--2830.

\bibitem{eurostat2020energy}
Eurostat, Energy consumption in households statistics,
  \url{https://ec.europa.eu/eurostat/statistics-explained/index.php/Energy_consumption_in_households#Energy_products_used_in_the_residential_sector},
  accessed 5/12/20 (June 2020).

\bibitem{ngeso2020fes}
{National Grid ESO}, Future energy scenarios 2020,
  \url{https://www.nationalgrideso.com/future-energy/future-energy-scenarios/fes-2020-documents},
  accessed 8/2/21 (2019).

\bibitem{ng2013etsy}
{National Grid}, Electricity ten year statement,
  \url{http://fes.nationalgrid.com/fes-document/fes-archives/}, accessed 8/2/21
  (2013).

\bibitem{ofgem2014electricity}
{OFGEM}, Electricity capacity assessment report 2014,
  \url{https://www.ofgem.gov.uk/} (2014).

\bibitem{ofgem2020existing}
OFGEM, Existing and future interconnector projects,
  \url{https://www.ofgem.gov.uk/electricity/transmission-networks/electricity-interconnectors},
  accessed 8/2/21 (2020).

\bibitem{ngeso2020data}
{National Grid ESO}, Data explorer,
  \url{https://www.nationalgrideso.com/balancing-data/data-finder-and-explorer},
  accessed 8/2/21 (2020).

\bibitem{ngeso2020winter}
{National Grid ESO}, Winter outlook,
  \url{https://www.nationalgrideso.com/research-publications/winter-outlook},
  accessed 8/2/21 (2021).

\bibitem{national2019security}
{National Grid ESO}, National electricity transmission system security and
  quality of supply standard (version 2.4),
  \url{https://www.nationalgrideso.com/document/141056/download}, accessed
  8/2/21 (2019).

\bibitem{beis2020dukes}
{UK Department of Business, Energy and Industrial Strategy (BEIS)}, Digest of
  {UK} energy statistics ({DUKES}) (2020).

\bibitem{ngg2020data}
{National Grid Gas}, Data and operations,
  \url{https://www.nationalgridgas.com/data-and-operations}, accessed 8/2/21
  (2020).

\bibitem{watson2019decarbonising}
S.~Watson, K.~J. Lomas, R.~A. Buswell, Decarbonising domestic heating: What is
  the peak {GB} demand?, Energy policy 126 (2019) 533--544.

\bibitem{staffell2012review}
I.~Staffell, D.~Brett, N.~Brandon, A.~Hawkes, A review of domestic heat pumps,
  Energy \& Environmental Science 5~(11) (2012) 9291--9306.

\bibitem{ccc2020sixth}
{Climate Change Committee}, The sixth carbon budget: The {UK}'s path to net
  zero, \url{https://www.theccc.org.uk/publication/sixth-carbon-budget/},
  accessed 8/2/21 (December 2020).

\bibitem{beis2019sub}
{UK Department of Business, Energy and Industrial Strategy (BEIS)},
  Sub-national electricity and gas consumption,
  \url{https://assets.publishing.service.gov.uk/government/uploads/system/uploads/attachment_data/file/853760/sub-national-electricity-and-gas-consumption-summary-report-2018.pdf},
  accessed 8/2/21 (December 2020).

\bibitem{zachary2014estimation}
S.~Zachary, C.~Dent, Estimation of joint distribution of demand and available
  renewables for generation adequacy assessment, arXiv preprint arXiv:1412.1786
  (2014).

\bibitem{sheehy2016impact}
S.~Sheehy, G.~Edwards, C.~J. Dent, B.~Kazemtabrizi, M.~Troffaes, S.~Tindemans,
  Impact of high wind penetration on variability of unserved energy in power
  system adequacy, in: 2016 International Conference on Probabilistic Methods
  Applied to Power Systems (PMAPS), IEEE, 2016, pp. 1--6.

\bibitem{ngeso2020ecr}
{National Grid ESO}, Electricity capacity report 2020,
  \url{https://www.emrdeliverybody.com/CM/Guidance.aspx}, accessed 8/2/21
  (2020).

\bibitem{thornton2016role}
H.~Thornton, B.~J. Hoskins, A.~Scaife, The role of temperature in the
  variability and extremes of electricity and gas demand in {Great Britain},
  Environmental Research Letters 11~(11) (2016) 114015.

\bibitem{beis2020letter}
{UK Department of Business, Energy and Industrial Strategy (BEIS)}, Capacity
  market auction parameters: letter from {BEIS to National Grid ESO [May 2019;
  July 2020]}, \url{https://www.gov.uk/business/uk-energy-security}, accessed
  16/2/21 (2020).

\end{thebibliography}

\end{document}